\documentclass[a4paper,11pt]{article}
\usepackage{jheppub} 

\usepackage{lineno}

\usepackage{amsmath}
\usepackage{amsfonts}
\usepackage{amssymb}
\usepackage{amstext} 
\usepackage{xfrac}
\usepackage{latexsym}
\usepackage{slashed}
\usepackage{array}
\usepackage{braket}
\usepackage{dsfont}
\usepackage[compat=1.0.0]{tikz-feynman}
\usetikzlibrary{shapes.geometric}
\usepackage{enumerate}
\usepackage{tabularx}
\usepackage{booktabs}
\usepackage{bm}
\usepackage{mathtools}
\usepackage{gensymb}
\usepackage{tensor}
\usepackage[table]{}
\usepackage{colortbl}

    \makeatletter
    \def\CT@@do@color{%
      \global\let\CT@do@color\relax
            \@tempdima\wd\z@
            \advance\@tempdima\@tempdimb
            \advance\@tempdima\@tempdimc
    \advance\@tempdimb\tabcolsep
    \advance\@tempdimc\tabcolsep
    \advance\@tempdima2\tabcolsep
            \kern-\@tempdimb
            \leaders\vrule
                    \hskip\@tempdima\@plus  1fill
            \kern-\@tempdimc
            \hskip-\wd\z@ \@plus -1fill }
    \makeatother

\usepackage{graphicx, rotating}
\usepackage{epstopdf}
\usepackage{epsfig}
\usepackage{caption}
\usepackage{subcaption}

\usepackage{lipsum}
\usepackage{comment}
\usepackage{scalerel}
\usepackage{multirow}
\usepackage{soul}
\usepackage{setspace}
\usepackage{xltabular}

\renewcommand{\arg}{\,\text{arg}\,}
\renewcommand{\Re}{\,\text{Re}\,}
\renewcommand{\Im}{\,\text{Im}\,}
\renewcommand{\mod}{\:\text{mod}\:}

\newcommand{\lagr}{\mathcal{L}}
\newcommand{\eps}{\epsilon}

\newcommand{\Oone}{\mathcal{O}(1)}

\newcommand{\CKM}{V_{\text{CKM}}}

\newcommand{\QCD}{_\text{QCD}}

\newcommand{\SUL}{SU(2)_{\text{L}}}

\newcommand{\FN}{\text{FN}}

\newcommand{\vphi}{v_{\Phi}}
\newcommand{\ma}{m_a}

\newcommand{\mrho}{m_\rho}
\newcommand{\Mp}{M_\text{Pl}}

\newcommand{\Z}{\mathbb{Z}}
\newcommand{\U}{\mathrm{U}}
\newcommand{\SU}{\mathrm{SU}}

\newcommand{\eminus}{\vcenter{\hbox{\scalebox{0.6}[1]{$ - $}}}}	

\usepackage{color}

\title{\boldmath Froggatt-Nielsen ALP}

\author{Admir Greljo,}%
\author{Aleks Smolkovi\v c,}%
\author{Alessandro Valenti}
\affiliation{%
 Department of Physics, University of Basel,\\Klingelbergstrasse 82, CH-4056 Basel, Switzerland}%
\emailAdd{admir.greljo@unibas.ch}
\emailAdd{aleks.smolkovic@unibas.ch}
\emailAdd{alessandro.valenti@unibas.ch}

\abstract{The Froggatt-Nielsen (FN) mechanism, a prominent framework for explaining the observed flavor hierarchies, generically predicts the existence of an axion-like particle (ALP). This work examines a class of FN models based on $\Z_N$ discrete symmetries.
We chart the allowed parameter space from a set of theoretical considerations and construct explicit renormalizable completions with minimal field content necessary to generate consistent textures. 
We then conduct comprehensive phenomenological analyses of two particularly elegant $\Z_4$ and $\Z_8$ models, highlighting the interplay between the effects of the ALP and the associated UV fields. We find that the FN scale can be as low as a few TeV.
}

\begin{document}
\maketitle
\flushbottom

\section{Introduction}
\label{sec:intro}

The observed pattern of fermion masses and mixings cannot be explained within the Standard Model (SM). Despite being generated from Yukawa interactions with a single Higgs field, one particularly intriguing aspect is the presence of roughly two-order-of-magnitude mass hierarchies between consecutive generations of all three types of charged fermions: up quarks, down quarks, and charged leptons. Additionally, the CKM mixing matrix~\cite{Cabibbo:1963yz, Kobayashi:1973fv} is approximately a unit matrix, with hierarchies evident in its off-diagonal elements. In stark contrast, the neutrino sector displays a different behavior. Apart from the small overall mass scale, neutrino oscillations reveal large mixing angles and a seemingly anarchic flavor structure. The origin of generation hierarchies in the charged fermion sector, in contrast to the neutrino sector, necessitates an organizing principle beyond the SM.

The Froggatt-Nielsen (FN) mechanism~\cite{Froggatt:1978nt, Leurer:1992wg, Leurer:1993gy} is a prototypical example of how approximate flavor symmetries can be used to generate Yukawa textures. This framework introduces a global $\U(1)_{\FN}$ symmetry, with different generations assigned different integer charges and a symmetry-breaking spurion $\epsilon \ll 1$ carrying a unit charge. By judiciously selecting these charges, one can construct Yukawa matrices $Y^f_{ij}$ that exhibit a desired suppression factor $\epsilon^{n_{ij}}$. The simplicity of a FN model often conflicts with its ability to accurately fit the observed flavor parameters~\cite{Fedele:2020fvh, Cornella:2023zme}. Opting for simpler charge assignments, with a relatively small largest integer charge, results in increased variance in the $\mathcal{O}(1)$ ultraviolet (UV) parameters. Nonetheless, such scenarios might be preferable from a simplicity perspective, given the uncertainties in the distribution of marginal couplings in the UV.

A straightforward way to realize $\epsilon$ is through the spontaneous symmetry breaking of a perturbatively exact global $\U(1)_{\FN}$ symmetry. This involves a complex scalar field $\Phi$ (a SM gauge singlet) with charge 1, which acquires a vacuum expectation value (VEV) $\langle \Phi \rangle = v_\Phi / \sqrt{2}$. Additionally, this framework assumes a gap between $v_\Phi$ and the common UV scale $M$, which underlies the generation of the FN effective field theory (EFT) operators. Consequently, $\epsilon = v_\Phi / \sqrt{2}M$.

As a result of a spontaneously broken continuous global symmetry, a Goldstone boson emerges in the spectrum, dominating infrared (IR) physics. This boson is not exactly massless due to the chiral anomaly with QCD, where non-perturbative instanton effects endow it with a mass, making it the QCD axion~\cite{Peccei:1977hh, Wilczek:1977pj, Weinberg:1977ma}. The QCD axion, proposed to address the strong CP problem, has recently been the focus of extensive theoretical~\cite{DiLuzio:2020wdo} and experimental studies~\cite{Irastorza:2018dyq}. The concept of the \textit{axiflavon}~\cite{Calibbi:2016hwq} or \textit{flaxion}~\cite{Ema:2016ops} is particularly compelling, as it simultaneously addresses two structural issues of the Standard Model: the flavor puzzle and the strong CP problem. This leads to intriguing phenomenology, characterized by the interplay between astrophysical and flavor physics constraints. Notably, the flavor constraints from $K^+ \to \pi^+ a$ impose the most stringent limits on the FN breaking scale, requiring $v_\Phi \gtrsim \mathcal{O}(10^{12})$\,GeV, as detailed in~\cite{MartinCamalich:2020dfe}.

The $\U(1)$ symmetry can be explicitly broken in various ways, which would undermine its solution to the strong CP problem. The most problematic and perhaps unavoidable breaking arises from Planck-scale suppressed operators, leading to the axion quality problem~\cite{Georgi:1981pu, Dine:1986bg, Barr:1992qq, Kamionkowski:1992mf, Holman:1992us}. The primary objective of this work is to explore a broader parameter space for the FN axion-like particle (ALP) beyond the axiflavon limit while delegating the resolution of the strong CP problem to another source. Explicit breaking can arise from operators of different dimensions. For instance, consider a soft breaking by a $m^2 \phi^2$ operator, which predicts the ALP mass $m_a \sim m$. A consistency condition is required to avoid disrupting the FN EFT power counting and maintaining the desired textures: $m_a \lesssim v_\Phi$.

A systematic approach to exploring this terrain involves considering perturbatively exact discrete subgroups. In this work, we focus on $\Z_N$. We find that the simplest consistent model in non-supersymmetric theories has $N = 4$ and already fits the observed flavor parameters at the level of recently proposed $\SU(2)$ gauged flavor models~\cite{Greljo:2023bix, Greljo:2024zrj} or flavor deconstruction examples~\cite{Davighi:2023iks, Greljo:2024ovt}. Increasing $N$ allows for a finer fit to the observed fermion masses and mixings. For instance, we demonstrate that $\Z_8$ can realize the same Yukawa structure as the recently proposed $\U(2)_{q+e^c+u^c} \times \Z_2$ symmetry~\cite{Antusch:2023shi}. 

Imposing a discrete symmetry instead of a continuous one eliminates the presence of a Goldstone boson in the theory. Specifically, for $\Z_4$, no light states are expected, with $m_a \sim v_\Phi$ predicted from the $\Phi^4$ term in the scalar potential. A light ALP is anticipated for $N > 4$, with its mass generated by a higher-dimensional operator. In fact, the same scale $M$ responsible for the Yukawa textures can also generate the ALP mass, as illustrated later by the \textit{wheel} models (see Fig.~\ref{fig:s2:wheel}). A natural outcome of this class of models is an FN ALP with mass $\gtrsim$ GeV, where the decay $K^+ \to \pi^+ a$ is kinematically forbidden, as illustrated by the $\Z_8$ model in Section~\ref{sec:Z8}.

This raises an intriguing question: what is the lowest FN scale that remains compatible with experimental constraints? This is a quantitative question with an interesting interplay between ALP-induced effects and UV contributions. In this paper, we conduct a thorough phenomenological study performing precision calculations in explicit renormalizable models, completing the FN $\Z_4$ and $\Z_8$ structures with minimalistic ingredients. The philosophy here is that rather than estimating UV contributions through FN EFT power counting, we introduce only the necessary UV field content to realize the FN mechanism and study its irreducible effects. As we demonstrate for the chosen models, the limit on $v_\Phi$ is weaker than the one estimated using the FN EFT. In contrast to the axiflavon case, this scenario motivates direct searches for heavy new physics at future colliders.

The paper is organized as follows. In Section~\ref{sec:theory}, we provide a theoretical discussion of $\Z_N$ models, charting the parameter space of the FN ALP. Section~\ref{sec:Z4} focuses on constructing a concrete model based on $\Z_4$ symmetry, fitting the SM flavor parameters, and presenting a comprehensive survey of experimental bounds. In Section~\ref{sec:Z8}, we construct and examine an explicit $\Z_8$ model. Finally, in Section~\ref{sec:conc}, we conclude our study. Technical details are discussed in the Appendices.

\section{$\Z_N$ Froggatt-Nielsen}
\label{sec:theory}

This section introduces a class of discrete FN models based on $\Z_N$ symmetries and provides a theoretical discussion of the ALP parameter space. 

\subsubsection*{Yukawa textures} 

Consider an EFT comprised of the SM matter content supplemented by a singlet complex scalar field $\Phi$. Both $\Phi$ and the SM fermions are assumed to be charged under a global, perturbatively exact\footnote{For realistic choices of the chiral fermion charges, the $\Z_N$ symmetry is typically anomalous. For recent studies of anomaly free FN models see Refs.~\cite{Alonso:2018bcg, Rathsman:2019wyk, Smolkovic:2019jow, Bonnefoy:2019lsn, Rathsman:2023cic}.} $\Z_N$ symmetry, with $[\Phi] =1$. The mechanism generating hierarchies in the SM Yukawas is of a Froggatt-Nielsen type:
\begin{align}
    -\lagr \supset \sum_{f,F} \left[  x_{ij}^f \left(\frac{\Phi}{M}\right)^{n_{ij}^f} +  x_{ij} ^{f\prime} \left(\frac{\Phi^*}{M}\right)^{n_{ij}^{ f \prime}}  \right] \bar F_{L,i} H  f_{R,j} +\{\widetilde{H}\}+\text{h.c.},
    \label{eq:s2:YukLagr}
\end{align}
where $x_{ij}^f, x_{ij} ^{f \prime}$ are a priori $\Oone$ couplings, and $n_{ij}^f$ and $n_{ij}^{ f \prime}$ are determined by the different $\Z_N$ charges of the SM fields. Here, we assume a common mass scale $M$ that generates the effective operators.
The term $\{\widetilde{H}\}$ is a shorthand notation for the interactions involving $\widetilde{H} \equiv \varepsilon H^*$, while $f$ and $F$ denote SM $\SUL$ singlets and doublets, respectively. $H$ is uncharged under $\Z_N$, $[H]=0$.

It is instructive to think of $\Z_N \subset \U(1)$. The $\U(1)$-breaking but $\Z_N$-preserving interactions involving $\Phi^*$ ($\Phi$) represent an important difference compared to the ``standard'' $\U(1)$ case. The $\Z_N$, but not the $\U(1)$, allows the replacement $\Phi^{n_{ij}^f} \rightarrow (\Phi^*)^{N-n_{ij}^f}$ in Eq.~\eqref{eq:s2:YukLagr}, since $[\Phi]^* = -[\Phi] \equiv (N-[\Phi]) \, \mod N$. To avoid spoiling the hierarchies generated exclusively by $\braket{\Phi}$, it is required that $n_{ij}^f \leq \left\lfloor \frac{N}{2} \right\rfloor$. Therefore, special care must be taken with charge assignments to ensure the desired textures. 
A sufficient condition is that chiral fermions of the SM are assigned charges from the set $\{0, \dots ,\left\lfloor \frac{N}{2} \right\rfloor \}$. For example, the smallest group that generates hierarchies between consecutive families by setting each family distinct charges $\{2,1,0\}$ is $\Z_4$ instead of $\Z_3$.\footnote{In this work, we focus on non-supersymmetric theories. Holomorphy, however, would make the $\Z_3$ symmetry eligible. For a study of $\Z_N$ FN models in supersymmetric context, see \cite{Higaki:2019ojq}.} Therefore, we will focus only on models with $N \geq 4$.

The numerical value for the spurion $\epsilon = \langle \Phi \rangle / \sqrt{2}M$ depends non-trivially on $N$ and the specific details of the model. However, a simple estimate can be made by requiring the largest hierarchy to be saturated by $\epsilon^{\left\lfloor \frac{N}{2} \right\rfloor} \sim y_e/y_t$, where the ratio of the electron to top quark Yukawa couplings $y_e/y_t \sim 10^{-6}$. Since $\epsilon \xrightarrow[]{N \rightarrow \infty} 1 $, large $N$ setups are disfavored not only from the simplicity perspective but also due to concerns regarding the consistency of the FN EFT expansion.

\begin{figure}[!t]
    \centering
    \includegraphics[width=\textwidth]{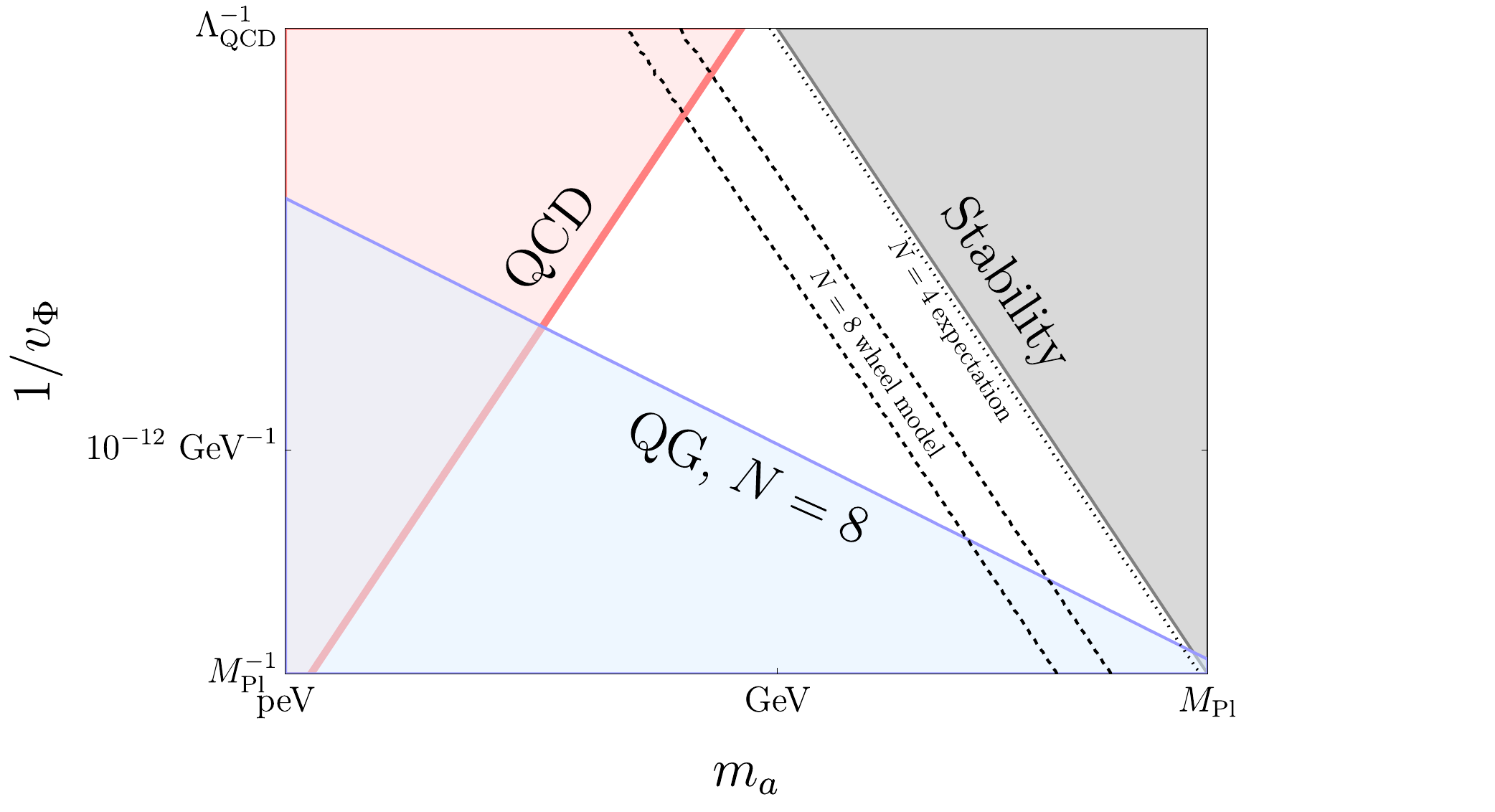}
    \caption{Theoretical constraints on the FN ALP, presented in the $(m_a, 1/\vphi)$ plane, arising from Eqs.~\eqref{eq:s2:massesRange}, \eqref{eq:s2:QGbound}, and \eqref{eq:s2:QCDbound}, with $\lambda=0.5$. The plots have been drawn for $N=4$ and $8$; in particular, the thickness of the darker gray and red lines represents the variability of the boundary from $N=8$ (less constraining) to $N=4$ (more constraining). The expectation for a representative $N=4$ model with $\lambda'_4 = \lambda/4$ is depicted with a dotted line, while the band enclosed by the dashed lines corresponds to the expectation around Eq.~\eqref{eq:s2:wheelma} for a $N=8$ ``wheel'' model. For details, see Section~\ref{sec:theory}.}
    \label{fig:s2:theoryPlot}
\end{figure}

\subsubsection*{Charting the ALP} 

Theoretical constraints that define the ALP's parameter space of interest are illustrated in Fig.~\ref{fig:s2:theoryPlot}. This figure presents various constraints, which will be discussed below, in the plane of ALP mass versus the inverse FN scale. The most general potential compatible with the $\Z_N$ symmetry reads
\begin{align}
    \begin{aligned}
    V(\Phi) &= - m^2 |\Phi|^2 +\frac{1}{2}\lambda |\Phi|^4 -  \frac{1}{4}\frac{\lambda_N'}{M_\Phi^{N-4}}  \left[\Phi^N + (\Phi^*) ^N \right] + \dots,\\
    \end{aligned}
    \label{eq:s2:potential}
\end{align}
where without loss of generality, we take $\lambda'_N \in \mathds{R}_{>0}$. For simplicity, we are neglecting any portals with the SM Higgs doublet and further higher-dimensional operators, which will lead to subleading corrections in the following discussion. 

Let us employ a convenient parametrization $\Phi = \left(\frac{v_\Phi+\rho}{\sqrt{2}}\right) e^{ia/\vphi}$, where $\rho$ and $a$ represent the radial and angular (ALP) modes, respectively, with $\langle \Phi \rangle = v_\Phi / \sqrt{2}$. The SM Yukawa hierarchies are explained by a small parameter $\eps \equiv v_\Phi/(\sqrt{2} M)$ after expanding Eq.~\eqref{eq:s2:YukLagr}. Since, in general, $M$ and $M_\Phi$ may differ, we define an effective coupling
\begin{align}
    \lambda_{4,\text{eff}}' =  \lambda_N ' \left(\frac{M}{M_\Phi} \right)^{N-4} \eps^{N-4}\,.
\end{align}
The minima of the potential in Eq.~\eqref{eq:s2:potential} are set at $\vphi^2 = 2m^2/(\lambda-N\lambda_{4,\text{eff}}'/4)$ and $\braket{a/\vphi}=2\pi k  /N$, $k =1, 2,\dots N$. The masses of the radial and angular modes read
\begin{align}
    m_{\rho} ^2 = \left(\lambda - \frac{1}{8} N(N-2)\lambda_{4,\text{eff}}' \right) \vphi^2 \quad \text{and}\quad  
    m_{a} ^2 =  \frac{1}{8} N^2 \lambda_{4,\text{eff}}'  \vphi^2 \,.
    \label{eq:s2:masses}
\end{align}
They satisfy a sum rule:
\begin{align}
    m_{\rho} ^2 + \left(\frac{N-2}{N}\right)  m_{a} ^2 = \lambda \vphi ^2\,.
    \label{eq:s2:massSumRule}
\end{align}
This, combined with the stability conditions $ \mrho^2 \geq 0$ and $\ma^2 \geq 0$, provides an admissible range for the $\rho$ and $a$ masses:\footnote{This also holds when taking into account possible subleading corrections to Eq.~\eqref{eq:s2:potential} by defining effective couplings $\lambda_\text{eff} = \lambda (1+ \mathcal{O} (\eps^2))$ and $\lambda_\text{4, eff} '  = \lambda_N ' \left(M/M_\Phi \right)^{N-4} \eps^{N-4} (1+ \mathcal{O} (\eps^{2}))$. Higher-order harmonics cannot, however, be described in this simple picture. They are suppressed by a factor of $\epsilon^N$ compared to the leading term and can thus be safely neglected.}
\begin{align}
\phantom{\qquad \textbf{[Stability condition]} }
\begin{aligned}
 0\leq \mrho^2&\leq \lambda \vphi^2, \\ 
    0\leq \ma^2&\leq\left(\frac{N}{N-2} \right) \lambda \vphi^2. 
\end{aligned}\qquad \textbf{[Stability condition]} 
    \label{eq:s2:massesRange}
\end{align}
The excluded region is shown in gray in Fig.~\ref{fig:s2:theoryPlot}. With marginal coupling $\lambda$ of $\mathcal{O}(1)$, we generally expect $m_\rho \sim v_\Phi$. The $N=4$ case is unique because the $\U(1)$-breaking operator is renormalizable, resulting in $m_a \sim  v_\Phi$ for $\lambda'_N \sim \Oone$ and, consequently, no ALP.\footnote{The stability of the $\Z_4$ potential requires $\lambda_4'<\lambda$. In the phenomenology section, we will also consider $\lambda_N' \ll 1$, in which case $a$ is an ALP.} However, for $N > 4$, there is an ALP with mass $m_a \ll v_\Phi$.

\begin{figure}[!t]
	\centering
		\resizebox{5.0cm}{!}{
			\begin{tikzpicture}
    \draw[thick] (0,0) circle (1.5);
    \foreach \angle in {45,90,...,315} {
    \draw[dashed, thick] (\angle:1.5) -- ++(\angle:0.75);
    \draw [dotted,thick,domain=-20:20] plot ({2*cos(\x)}, {2*sin(\x)});
    }
    \node at (-2.5,0) {$\Phi$};
    \node at (-1.79,1.79) {$\Phi$};
    \node at (0,2.5) {$\Phi$};
    \node at (1.79,1.79) {$\Phi$};
    \node at (1.79,-1.79) {$\Phi$};
    \node at (0, -2.5) {$\Phi$};
    \node at (-1.79,-1.79) {$\Phi$};
    \node at (-1.65,0.8) {$\Psi_1$};
    \node at (-0.8, 1.65) {$\Psi_2$};
    \node at (0.8, 1.65) {$\Psi_3$};
    \node at (0.8, - 1.7) {$\Psi_{N-2}$};
    \node at (-0.8, -1.65) {$\Psi_{N-1}$};
    \node at (-1.65,- 0.8) {$\Psi_N$};
\end{tikzpicture}
		}
		\caption{The ``wheel'' diagram generating the $\U(1)_{\FN}$ breaking term $\Phi^N + (\Phi^*)^N$. For details see Section~\ref{sec:theory}.}
		\label{fig:s2:wheel}
	\end{figure}
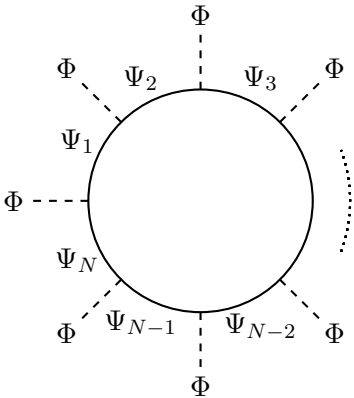

Let us sketch a renormalizable UV model example that generates $\lambda_N'$ for $N>4$ and predicts the ALP mass. Any completion with only a minimal set of vector-like fermions (VLFs) required to generate Yukawa operators in Eq.~\eqref{eq:s2:YukLagr} with desired textures typically leads to an accidental $\U(1)_{\FN}$ symmetry when these fields are integrated out, thus failing to generate $\lambda_N'$. An elegant mechanism to generate $\lambda_N'$ would be to consider a \emph{full} set of VLFs $\Psi_k$ with $\Z_N$ charges $[\Psi_k] = k$ and couplings $\sum_k y_k \, \Phi \, \bar \Psi _{L,k+1} \Psi_{R,k}$. The ``wheel'' diagram in Fig.~\ref{fig:s2:wheel} predicts
\begin{align}
     \frac{\lambda_N '}{M_\Phi^{N-4}} \sim \frac{1}{M_\Psi ^{N-4}}\frac{\prod_k y_k}{16 \pi^2}\,,
     \label{eq:s2:wheelCoeff}
\end{align}
under the assumption of approximately degenerate $\Psi_i$ masses.\footnote{A straightforward mechanism to achieve this would be to introduce an additional real scalar field $\chi$ such that $(\chi, \Phi, \Psi_{R,k})$ are odd under a new \(\mathbb{Z}_2\) symmetry. When $\chi$ acquires a non-zero vacuum expectation value, the interactions $\chi \bar{\Psi}_{L,k} \Psi_{R,k}$ result in comparable masses among all VLFs.} 

Incidentally, a subset of these VLFs, with the appropriate couplings, is sufficient to generate the Yukawa operators in Eq.~\eqref{eq:s2:YukLagr}. Therefore, in this class of models,
\begin{align}
M\sim M_\Phi \sim M_\Psi.
\end{align}
This, finally, allows us to predict the ALP mass using Eq.~\eqref{eq:s2:masses},
\begin{equation}\label{eq:s2:wheelma}
    \frac{m_a}{v_\Phi} \sim \frac{1}{8\pi} N \epsilon^{\frac{N}{2} -2} \sqrt{\prod_k y_k}~,
\end{equation}
where we take an estimate for $\epsilon \sim (y_e / y_t)^{\left\lfloor \frac{N}{2} \right\rfloor^{-1}}$, as discussed earlier. The prediction for the $N = 8$ case is shown in Fig.~\ref{fig:s2:theoryPlot} as a region within the two black dashed lines which represent the variation $0.001 < \sqrt{\prod_k y_k} < 0.2$.\footnote{Choosing $y_k$ from a flat distribution in the $(0,1)$ range, there is $90\%$ probability for $\sqrt{\prod_k y_k}$ to be above (below) 0.001 (0.2).}

This mechanism might, however, not be present at all, and the $\U(1)_{\FN}$-breaking term in Eq.~\eqref{eq:s2:potential} could be generated by UV physics at scales $M_\Phi$ significantly larger than $M$. In the most extreme scenario, the $\U(1)_{\FN}$ symmetry is broken solely by quantum gravity effects $M_\Phi \sim \Mp$, which are expected to break any global symmetry~\cite{Palti:2019pca}.\footnote{In this scenario, we assume that $\mathbb{Z}_N$ symmetry is protected, possibly through an appropriate gauging mechanism \cite{Krauss:1988zc, Banks:1989ag, Preskill:1990bm, Ibanez:1991hv}.} If this were the case, the coupling $\lambda_{4, \text{eff}}'$ would be lower bounded as $\lambda_{4, \text{eff}}' \gtrsim  16\pi^2 (M/\Mp)^{N-4} \eps^{N-4}$, barring any fine-tuning on Planck-suppressed operators.\footnote{Here $\Mp$ denotes the reduced Planck mass, $\Mp\simeq 2.4 \times  10^{18}$ GeV.} This leads, in turn, to a lower bound on the ALP mass:
\begin{align}
   \phantom{ \qquad \textbf{[QG condition]}} m_a ^2 \gtrsim  8\pi^2 \left(\frac{N^2}{ 2^{N/2}} \right) \left(\frac{\vphi}{\Mp} \right)^{N-4} \vphi^2\,. \qquad \textbf{[QG condition]} 
   \label{eq:s2:QGbound}
\end{align}
This lower bound is analogous in spirit to the constraint on the dimension of PQ-breaking operators encountered in the well-known QCD axion quality problem~\cite{Georgi:1981pu, Dine:1986bg, Barr:1992qq, Kamionkowski:1992mf, Holman:1992us}. The ``excluded" region for $N=8$ is shown in blue in Fig.~\ref{fig:s2:theoryPlot}.

Finally, one should not forget QCD itself and its contribution to the axion mass, which is approximately given by $(\ma^2)\QCD \sim m_\pi ^2 f_\pi ^2 c_a^2/ v_\Phi^2$, where $c_a$ is the anomaly coefficient entering $(c_a g^2_s/ 32\pi^2) G\widetilde{G}$, and we approximate $c_a \sim N$. As the physical mass of the ALP is given by the sum of this contribution and the one of Eq.~\eqref{eq:s2:masses}, any value of $m_a$ lower than that of the single contributions requires a non-trivial cancellation. Avoiding this tuning can then be stated as
\begin{align}
    \phantom{\qquad \textbf{[QCD condition]}} \ma \gtrsim  (\ma) \QCD. \qquad \textbf{[QCD condition]}
    \label{eq:s2:QCDbound}
\end{align}
The red region shown in Fig.~\ref{fig:s2:theoryPlot} does not satisfy this condition.\footnote{In the presence of multiple ALPs, this condition applies only to the linear combination entering the QCD potential. Hence, there might be additional lighter fields~\cite{Gavela:2023tzu}.}

To conclude, Eqs.~\eqref{eq:s2:massesRange}, \eqref{eq:s2:QGbound}, and \eqref{eq:s2:QCDbound} together define an acceptable region for $m_a$ and $v_\Phi$ that a $\Z_N$ FN model should respect. This corresponds to the white region in Fig.~\ref{fig:s2:theoryPlot}, where we also show model expectations for \(N=\{4, 8\}\) as discussed above.

\begin{table}[t]
\centering
\renewcommand{\arraystretch}{1.5}
\begin{tabular}{|c|c|c|c|c|}\hline
\rowcolor{black!15}[] Fields & $\SU(3)_C$ & $\SU(2)_L$ & $\U(1)_Y$ & $\Z_4$ \\ 
\hline
$q_{1,2,3}$ & $\textbf{3}$ & $\textbf{2}$ & $\frac{1}{6}$ & $\{2,1,0\}$ \\
$\ell_i$ & $\textbf{1}$ & $\textbf{2}$ & $\eminus\frac{1}{2}$ & $0$ \\
\hline
$e_{1,2,3}$ & $\textbf{1}$ & $\textbf{1}$ & $\eminus1$ & $\{\eminus2,\eminus1,0\}$ \\
$u_i$ & $\textbf{3}$ & $\textbf{1}$ & $\frac{2}{3}$ & $0$ \\
$d_i$ & $\textbf{3}$ & $\textbf{1}$ & $\eminus\frac{1}{3}$ & $0$ \\
\hline
$Q^{a}_2$ & $\textbf{3}$ & $\textbf{2}$ & $\frac{1}{6}$ & $0$ \\
$Q_1$ & $\textbf{3}$ & $\textbf{2}$ & $\frac{1}{6}$ & $1$ \\
$E^{a}_2$ & $\textbf{1}$ & $\textbf{1}$ & $\eminus1$ & $0$ \\
$E_1$ & $\textbf{1}$ & $\textbf{1}$ & $\eminus1$ & $1$ \\
\hline
$H$ & $\textbf{1}$ & $\textbf{2}$ & $\frac{1}{2}$ & $0$ \\
$\Phi$ & $\textbf{1}$ & $\textbf{1}$ & $0$ & $1$ \\ \hline
\end{tabular}
\caption{Field content of the UV model as described in Section~\ref{sec:Z4}. The rows in this table are grouped into four categories. The first and second represent left and right chiral fermions, respectively. The third consists of vector-like fermions, and the last comprises scalars. Flavor indices are denoted as $i=1,2,3$ and $a=1,2$. See Section~\ref{sec:z4setup} for details.}
\label{tab:z4fields}
\end{table}

\section{The minimal model: $\Z_4$}
\label{sec:Z4}

This section constructs and thoroughly analyses an explicit $\Z_4$ FN model. We begin in Section~\ref{sec:z4setup} by defining a concrete renormalizable model that realizes the $\Z_4$ FN texture, and we examine the EFT structure it generates. Following this, Section~\ref{sec:z4observables} offers a comprehensive analysis of the various observables predicted by the model. Section~\ref{sec:z4results} collects constraints in $(m_a, \vphi^{-1})$ plane and discusses their broader implications.

\subsection{Setup}
\label{sec:z4setup}

The $\Z_4$ symmetry offers the simplest realization capable of generating the SM fermion hierarchies. It features a renormalizable potential given by the $N\rightarrow4$ limit of Eq.~\eqref{eq:s2:potential}, ensuring no massless spin-0 states exist. Therefore, this case can be considered minimal in terms of UV matter content, as it only requires completing the Yukawa operators in Eq.~\eqref{eq:s2:YukLagr}.

\paragraph{UV model.}
The field content of a particular model incarnating this idea is listed in Table~\ref{tab:z4fields}. Apart from $\Phi$ and the SM fields, the model includes only the addition of vector-like quarks $Q_1, Q_2^{a=1,2}$ and vector-like leptons $E_1, E_2^{a=1,2}$.\footnote{Alternatively, an even more minimal completion involves replacing VLFs with three additional Higgs doublets carrying $\Z_4$ charges of $\{1, 2, 3\}$. Their masses are $\mathcal{O}(M)$, and they do not develop a VEV.} The non-trivial FN charges under the $\Z_4$ symmetry are $[Q_1] = -[E_1] = 1$. The only charged SM fields are $q$ and $e$, with $[q_1] = -[e_1] = 2$ and $[q_2] = -[e_2] = 1$.\footnote{This charge assignment is inspired by the recently proposed $\U(2)_{q+e}$ flavor symmetry~\cite{Antusch:2023shi}. Charging under the flavor group only left-handed quarks and right-handed leptons produces the observed mass hierarchies of electrically charged fermions and the mixing hierarchies of left-handed quarks while maintaining an anarchic flavor structure for neutrinos, ensuring large neutrino mixing. Our $\Z_4$ is the smallest flavor symmetry resulting in similar Yukawa textures as predicted by $\U(2)_{q+e}$ of~\cite{Antusch:2023shi}.} All the other fields are taken as $\Z_4$-neutral, except $\Phi$. The Yukawa sector of the most general renormalizable Lagrangian compatible with this symmetry reads\footnote{Note that Eq.~\eqref{eq:s3:L_VLFs} respects an accidental $\U(1)_\FN$. If we would have chosen a $\Z_3$ symmetry instead of $\Z_4$, additional couplings, such as $\overline{q}_1 \Phi^* Q^a_{2}$, would appear, violating $\U(1)_\FN$. This would disrupt the desired Yukawa textures, as discussed below Eq.~\eqref{eq:s2:YukLagr}.}
\begin{equation}
\label{eq:s3:L_VLFs}
\begin{split}
    \mathcal{L}_\text{UV}&\supset x_1^q \Phi \overline{q}_1  Q_1 + x_{12}^{q a} \Phi \overline{Q}_1 Q_2 ^a + x_2^{q a} \Phi \overline{q}_2  Q_2^a - y_{j}^{d a} \overline{Q}_2^a H d_j - y_{j}^{u a} \overline{Q}_2^a \tilde{H} u_j \\
    & +x_1^e \Phi \overline{E}_1  e_1 + x_{12}^{e a} \Phi \overline{E}_2^a E_1 +x_2^{e a} \Phi \overline{E}_2^{a} e_2 - y_i^{ea} \overline{\ell}_i H E_2^{a} \\
    & -z_j^d \bar{q}_3 H d_j - z_j^u \bar{q}_3 \tilde{H} u_j - z_i^e \bar{\ell}_i H e_3 + \text{h.c.}\,.
\end{split}
\end{equation}
Flavor symmetries of the kinetic terms are used to perform field redefinitions and eliminate mass mixings between chiral and vector-like fermions of the same representation without any loss of generality. The specific basis and parameterization used are detailed in \ref{sec:appUV}.

\begin{figure}[!t]
    \centering
    \includegraphics[width=\textwidth]{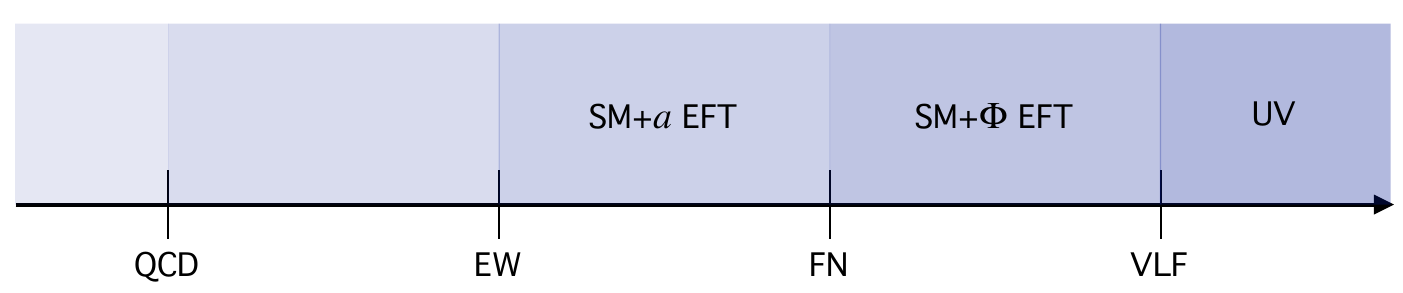}
    \caption{Illustration of the hierarchical scales within the model and the corresponding layers of effective field theory descriptions.}
    \label{fig:scales}
\end{figure}

\paragraph{Tower of EFTs.} A systematic analysis of this model involves constructing a tower of effective field theories, as illustrated in Fig.~\ref{fig:scales}. In the first step, the VLF states are integrated out, matching to the SM+$\Phi$ EFT. The resulting set of local higher-dimensional operators can be categorized into two broad classes based on their involvement of the $\Phi$ field. First are operators that include the $\Phi$ field. A subset of these include Eq.~\eqref{eq:s2:YukLagr}, which will generate the SM Yukawa couplings at the FN scale $v_\Phi$ upon matching to the next EFT, the SM+$a$ EFT. This class of operators also generates $a$ couplings. As already discussed, $m_a$ derived in Eq.~\eqref{eq:s2:masses} is governed by $\lambda_N'$. Although the natural expectation $ \lambda_N' \sim 1 $ predicts $m_a \sim v_\Phi $, in this section, we treat $ \lambda_N' $ as a free parameter that can be arbitrarily small.\footnote{It is technically natural to take $\lambda_4' \ll 1$, though this opposes the spirit of our narrative.} Our goal is to explore a broader parameter space of ALP masses, deriving general lessons that relate to $N > 4$ models where ALP is expected to be light. Second, another class of operators exists that does not involve the $\Phi$ field. These operators violate the accidental flavor symmetries of the SM and induce significant UV contributions to rare and forbidden SM transitions, as we will see in a moment.

\paragraph{Generating the SM Yukawas.} Integrating out the VLF at the tree-level leads to the effective Yukawa Lagrangian of Eq.~\eqref{eq:s2:YukLagr} with $x_{ij} ^{f\prime} = 0$ and
\begin{equation}
\label{eq:s3:nij}
    n_{ij} ^{d} = n_{ij} ^{u} =(n_{ij} ^{e})^t=
    \begin{pmatrix}
        2 & 2 & 2\\
        1 & 1 & 1\\
        0 & 0 & 0
    \end{pmatrix}.
\end{equation}
Representative Feynman diagrams are shown in Fig.~\ref{fig:diag}. The matching of $x_{ij} ^f$ to the UV parameters in Eq.~\eqref{eq:s3:L_VLFs} is provided in Appendix~\ref{app:paramsMatching}, together with an explicit benchmark point in the parameter space correctly reproducing the SM fermion masses and mixings, see Eqs.~\eqref{eq:app1:UVbenchQuarks} and \eqref{eq:app1:UVbenchLeptons}. Interestingly, their values can be correctly predicted with most of the UV couplings in Eq.~\eqref{eq:s3:L_VLFs} within a factor of 10 from each other. This stems from the fact that the light SM Yukawas are expressed as products of many UV couplings, which can naturally contribute to the hierarchy if these are somewhat small, say $\mathcal{O}(0.3)$.\footnote{A similar effect is observed in Fig.~3 of \cite{Greljo:2024zrj}.} The only exceptions are $y_b$ and $y_\tau$, which are here not dynamically generated, and hence we require them to be of $\mathcal{O}(10^{-2})$.

\begin{figure}[!t]
    \centering
    \includegraphics[width=\textwidth]{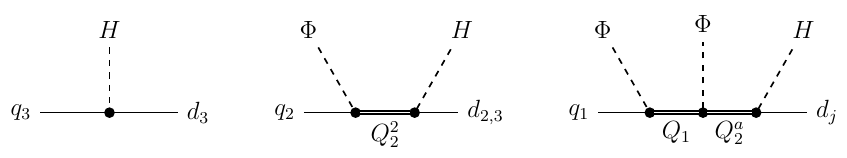}
\caption{Leading diagrams in the $\Z_4$ model generating effective down-quark Yukawa operators in Eq.~\eqref{eq:s2:YukLagr}, and analogously for up quarks and charged leptons. For further details, refer to Section~\ref{sec:z4setup}.}
    \label{fig:diag}
\end{figure}

\paragraph{ALP couplings.} For the upcoming flavor phenomenology discussion, we express the couplings of the radial and angular modes of $\Phi$ in the SM fermion mass basis as
\begin{equation}
\label{eq:s3:rhoa_couplings}
    \mathcal{L} \supset - \sum_{f=u,d,e}  c_{ij}^f \bar{f}_{Li} \left(\rho+ia\right) f_{Rj} + \text{h.c.} + \text{h.o.} \,,
\end{equation}
where the coupling matrices can be generally expressed as
\begin{equation}
    c^f = \frac{1}{v_\Phi }\left[ \left(U_L^{f\dagger} Q^F U_L^f \right) \hat{m}^f-  \hat{m}^f \left(U_R^{f\dagger} Q^f U_R^f \right)\right]\,.
\end{equation}
Here $U_L$ and $U_R$ are the left- and right-handed fermion rotation matrices that diagonalize the corresponding mass terms, $\hat{m}$ are the diagonal fermion mass matrices, and $Q$ are diagonal matrices encoding the SM fermion FN charges. The derivation of this formula is presented in Appendix~\ref{sec:app:ALP}. In our concrete realization of the FN model, only the left-handed quarks are charged, and the down-type quark coupling matrix can be written as
\begin{equation}
    c^d =\frac{1}{v_\phi }\left(V_\mathrm{CKM}^{\dagger} Q^d V_\mathrm{CKM} \right) \hat{m}^d \,,
\end{equation}
where we assumed that the CKM matrix is fully given by $U_L^d = V_\mathrm{CKM}$. This is a sensible assumption as the bigger hierarchies in the up sector lead to much more suppressed mixing angles in $U_L^u$, as shown in App.~\ref{app:paramsMatching}. Interestingly, the down-quark flavor-changing neutral currents due to $a,\rho$ are then fully determined by the measured values of the down-quark masses and the CKM matrix, up to the overall normalization of $v_\Phi$. Using the Wolfenstein parameterization, we get
\begin{equation}
\label{eq:s3:cdmatrix}
   c^d = \frac{1}{v_\phi } \left(
\begin{array}{ccc}
 2 m_d & \lambda  m_s & A \lambda ^3 m_b (-2 i \eta +2 \rho -1) \\
 \lambda  m_d & m_s & A \lambda ^2 m_b \\
 A \lambda ^3 m_d (2 i \eta +2 \rho -1) & A \lambda ^2 m_s & A^2 \lambda ^4 m_b \\
\end{array}
\right)\,,
\end{equation}
where in each entry, we keep only the leading nontrivial term in $\lambda$. The $c^d_{33}$ entry is small as a result of the fact that both the left and right components of the third generation are uncharged under $\Z_4$. Hence, the coupling only arises due to a small mixing. This is a peculiarity of the $\Z_4$ symmetry, which does not hold for larger $N$ where one attempts to address the smallness of $y_b$ and $y_\tau$ dynamically. For the purposes of phenomenology, we also report the lowest order in which we get complex off-diagonal entries:
\begin{align}
\begin{split}
    v_\Phi c^d _{12} &= \lambda  m_s - \lambda^3 m_s/2 +\lambda^5 m_s  (A^2 (i\eta-\rho+1) -1/8) \,,\\
    v_\Phi c^d _{21} &= \lambda  m_d -\lambda^3 m_d/2 + \lambda^5  m_d ( A^2 (-i \eta -\rho+1) -1/8)\,,\\
    v_\Phi c^d _{23}&= A\lambda^2 m_b + A\lambda^4 m_b(-2i\eta +2\rho -1/2) \,,\\
    v_\Phi c^d _{32}&= A\lambda^2 m_s + A\lambda^4 m_s(2i\eta +2\rho -1/2)\,.
\end{split}
\end{align}
Similarly, for $c^u$ and $c^e$ we find 
\begin{align}
\begin{aligned}
    c^u &= \frac{1}{\vphi}
    \begin{pmatrix}
        2m_u & m_c \frac{m_u}{m_c} \frac{z_{u_2}}{z_{u_1}} & m_t \frac{m_u}{m_t}\frac{2 y_{u_2}z_{u_3} - y_{u_3}z_{u_2}}{y_{u_2} z_{u_1}} \\
        m_u \frac{m_u }{m_c} \frac{z_{u_2}}{z_{u_1}} & m_c & m_t \frac{m_c}{m_t} \frac{y_{u_3}}{y_{u_2}}\\
        m_u \frac{m_u }{m_t} \frac{2 y_{u_2}z_{u_3}^* - y_{u_3}z_{u_2}}{y_{u_2} z_{u_1}} & m_c \frac{m_c }{m_t} \frac{y_{u_3}}{y_{u_2}}
        & m_c \frac{m_c }{m_t} \frac{y_{u_3}^2}{y_{u_2}^2}
    \end{pmatrix},\\ 
    c^e &= \frac{1}{\vphi}
    \begin{pmatrix}
        2 m_e & m_e \frac{m_e }{m_\mu} \frac{z_{e_2}}{z_{e_1}} & m_e \frac{m_e}{m_\tau} \frac{2 y_{e_2}z_{e_3} ^* - y_{e_3}z_{e_2}}{y_{e_2} z_{e_1}} \\
        m_\mu \frac{m_e}{m_\mu}\frac{z_{e_2}}{z_{e_1}} & m_\mu & m_\mu \frac{m_\mu }{m_\tau} \frac{y_{e_3}}{y_{e_2}} \\
        m_\tau \frac{m_e}{m_\tau}  \frac{2 y_{e_2}z_{e_3} - y_{e_3}z_{e_2}}{y_{e_2} z_{e_1}}  &  m_\tau \frac{m_\mu }{m_\tau} \frac{y_{e_3}}{y_{e_2}}  & m_\mu \frac{m_\mu}{m_\tau} \frac{y_{e_3}^2}{y_{e_2}^2} 
    \end{pmatrix},
\end{aligned}
\label{eq:s3:cecuMat}
\end{align}
where $y_{u_i}, z_{u_i}, y_{e_i}, z_{e_i}$ are parameters that are not determined within the EFT but can be correlated to the UV parameters in Eq.~\eqref{eq:s3:L_VLFs} as shown in App.~\ref{app:paramsMatching}. The matrices have been written factorizing the mixing, which renders evident the fact that ALP-induced flavor violation is suppressed by the ratio of the fermion masses. Again, the couplings to the third generation are small due to the fact that these are not charged under $\Z_4$. We stress that $c^d$ of Eq.~\eqref{eq:s3:cdmatrix} features the same pattern upon recognizing that $\lambda \sim m_d/m_s, A\lambda^2 \sim m_s/m_b$ and $A\lambda^3 \sim m_d/m_b$.

\paragraph{$\Z_4$-symmetric UV operators.} Integrating out VLFs at the UV matching scale directly generates certain SMEFT operators that violate approximate flavor symmetries and are crucial for the phenomenological discussion. We refer to these as ``VLF contributions", and we will carefully study their interplay with the ALP contributions. Already at the tree level, the model generates
\begin{equation}
    (H^\dag i\overset{\text{\footnotesize$\leftrightarrow$}}{D}_\mu H)(\bar \ell_i\gamma^\mu \ell_j), \quad (H^\dag i\overset{\text{\footnotesize$\leftrightarrow$}}{D}_\mu H)(\bar d_i\gamma^\mu d_j), \quad (H^\dag i\overset{\text{\footnotesize$\leftrightarrow$}}{D}_\mu H)(\bar u_i\gamma^\mu u_j),
\end{equation}
with arbitrary flavor structure and Wilson coefficients $\sim M^{-2}$. These operators provide a leading contribution to $\Delta F = 1$ transitions such as $\epsilon'/\epsilon$, $\mu \to e$ conversion, $K_L \to \mu^+\mu^-$, etc. Additionally, under renormalization, they mix into \(\Delta F = 2\) operators.
The primary \(\Delta F = 2\) contributions, however, arise from four-fermion operators generated through one-loop matching at the UV scale. These are
\begin{equation}\label{eq:sec3:4fops}
(\bar d_{i}\gamma^\mu d_j)(\bar d_{k}\gamma_\mu d_{l}), \quad (\bar u_{i}\gamma^\mu u_j)(\bar u_{k}\gamma_\mu u_{l}),
\end{equation}
with Wilson coefficients $\sim (4 \pi M )^{-2}$. The flavor violation in Eq.~\eqref{eq:sec3:4fops} is expected to be of $\mathcal{O}(1)$ since our model does not impose any selection rules. This will result in stringent constraints from neutral meson mixing, most notably $\eps_K$.
Finally, lepton flavor violating four-fermion operators
\begin{equation}
    (\bar \ell_i \gamma ^\mu \ell_j) (\bar d_k \gamma _\mu d_l),\quad (\bar \ell_i \gamma ^\mu \ell_j) (\bar u_k \gamma _\mu u_l),
\end{equation}
are generated at the one-loop level and they represent the leading contribution to LFV meson decays, such as $K_L\to e \mu$.

In the following subsection, we will review all the relevant observables and provide precise details to support these statements.

\subsection{Observables}
\label{sec:z4observables}

The $\Z_4$ model provides a concrete framework to explore the interplay between the VLF and the ALP contributions. As discussed above, we treat the $\U(1)_{\FN}$-breaking parameter $\lambda'_N$ as a free parameter and consider a broad range of ALP masses, temporarily setting aside the expectation of $m_a \sim v_\Phi$ for $\lambda'_N \sim 1$. By allowing $\lambda'_N \ll 1$, we can explore the ALP phenomenology in a concrete model. As we will argue later, the results should remain qualitatively consistent across the entire class of FN ALP models. Our presentation in this subsection is organized by observables, highlighting contributions from each source.\footnote{For related phenomenological ALP studies see~\cite{Bjorkeroth:2018dzu,Bauer:2019gfk, MartinCamalich:2020dfe, Bauer:2021mvw, Bonilla:2022qgm, Cornella:2023kjq,Biekotter:2023mpd, Calibbi:2024rcm, DiLuzio:2024jip, Knapen:2023zgi}.}

\subsubsection*{Rare meson decays}

\emph{ALP.} In the $\U(1)_\FN $ limit, $ a $ becomes very light, making the ``invisible" meson decay $ K \rightarrow \pi a $ the most significant constraint on $\vphi$, followed by $ B \rightarrow K^{(*)} a $. The decay rate is given by
\begin{align}
\begin{aligned}
    \Gamma ( M_1 \rightarrow M_2 \, a ) = \frac{m_{M_1}^3}{64\pi}  \lambda^{1/2} \left(\frac{m_{M_2}}{m_{M_1}},   \frac{m_a}{m_{M_1}}\right)  & \left(\frac{1-m_{M_2}^2 /m_{M_1}^2}{m_{q_i} - m_{q_j}} \right) ^2\left|f_{M_1\rightarrow M_2} (m_a^2) \right|^2 \\
    & \times \left[|c^f_{ij}|^2 + |c^f_{ji}|^2 -2 \Re c^f _{ij} c^f _{ji}\right],
\end{aligned}
    \label{eq:s3:mesondecays}
\end{align}
where $\lambda(x,y)=(1-(x + y)^2)(1-(x-y)^2)$ and $q_i, q_j$ refer to quarks from the parton-level transition. The quantity $f_{M_1\rightarrow M_2} (m_a^2)$ is the form factor parameterizing the hadronic matrix element evaluated at $q^2 = m_a ^2$, where $\braket{M_2 (p_2)| \bar q_i q_j | M_1 (p_1)} = f_{M_1 \rightarrow M_2} (q^2) (m_{M_1}^2 - m_{M_2}^2)/(m_{q_i}-m_{q_j})$ with $q=p_1-p_2$. For these, we rely on their determinations as reported in Refs.~\cite{FlavourLatticeAveragingGroupFLAG:2021npn, Gubernari:2023puw, Parrott:2022rgu, Bailey:2015dka, Bouchard:2013eph,Horgan:2013hoa,Horgan:2015vla}. The constraints we consider include the combined E787 and E949 analysis $\text{Br}(K^+ \rightarrow \pi^+  E_\text{miss}) < 9.5 \times 10^{-11}$ \cite{E949:2007xyy}, the BaBar bound $\text{Br}(B \rightarrow K^* E_\text{miss}) < 1.0\times 10^{-4}$ \cite{BaBar:2013npw} and the recent Belle II measurement $\text{Br}(B ^+ \rightarrow K^+ \bar \nu \nu) = (2.3 \pm 0.7)\times 10^{-5}$ \cite{Belle-II:2023esi}.~\footnote{The bounds in the paper are taken at 95 $\%$ CL. The ones given at $90\%$ CL have been converted following \cite{Calibbi:2017uvl}, when possible.} For the latter, given the current $\approx 2.4\sigma$ tension between the SM prediction and the experimental value, we employ the $3\sigma$ bound on $\text{Br}(B\rightarrow K a)$ derived in \cite{Altmannshofer:2023hkn} (see also \cite{Bolton:2024egx}).
For $ m_a < m_{K,B} $, these observables provide the most significant bounds on the FN scale, with $\vphi \gtrsim 10^{11}$\,GeV for kaon decay and $\vphi \gtrsim 10^7$\,GeV for $B$ decay.

Importantly, these bounds apply only if the ALP behaves as an invisible final state. This requires that the distance traveled by the ALP after production and before an eventual decay exceeds the length of the detector, $l_a > l_\text{detector}$, where 
\begin{align}
    l _a  = \frac{c |p_a|}{m_a \Gamma_a}.
    \label{eq:s3:distTravelled}
\end{align}
In this expression, $ |p_a| $ represents the axion momentum in the laboratory frame, and $ \Gamma_a $ is its total decay width. We approximate $ l_\text{detector} \approx 1 $\,m and assume the decaying meson is at rest.\footnote{These considerations also apply to $ M_1 \rightarrow M_2 \, \rho $ decays for a very light $\rho$, with the replacements $m_a \rightarrow m_\rho$ and $c^f_{ij} \rightarrow -i c^f_{ij}$. This applies to all processes discussed in this section; hence, we will report explicit expressions involving only $a$, particularly since a light $\rho$ is less motivated.}

If the ALP can decay inside the detector, the signatures differ. Prominent constraints in our model involve leptonic and semi-leptonic decays of flavored mesons \cite{Cornella:2019uxs}. The rate for lepton-flavor-violating decays is straightforward and reads
\begin{align}
\begin{aligned}
    \Gamma\left(M \rightarrow \ell _i \bar \ell _j\right) = \frac{f_M ^2 m_M }{64\pi }& \frac{\lambda^{1/2} \left(\frac{m_{\ell_i}}{m_M},  \frac{m_{\ell_j}}{m_M} \right)}{(m_{q_k} + m_{q_l})^2} \frac{\left[ |c^q_{kl}|^2 + |c^q_{lk}| +2 \Re c^q_{kl} c^q_{lk}\right]}{\left(1-m_{a}^2/m_M^2 \right)^2 + m_a ^2 \Gamma_a ^2/m_M^4 } \\
    & \times \left[ \left( |c^e_{ij}|^2 + |c^e_{ji}|^2\right) \left( 1- \frac{m_{\ell_i} ^2 + m_{\ell_j}^2}{m_M^2} \right) +4 \Re c^e _{ij} c^e _{ji} \frac{ m_{\ell_i} m_{\ell_i}}{m_M^2} \right],
\end{aligned}
\label{eq:s3:leptonicMesonDecaysRate}
\end{align}
where $q=u,d$. Subscripts $q_k$ and $q_l$ refer to the underlying quark transition, $\Gamma_{a}$ is the total width of $a$ and $\braket{0|\bar q_k \gamma^5 q_l|M} = f_M m_ M^2/(m_{q_k} + m_{q_l})$. We take the $95\%$\,CL bounds on the branching ratios from Table~2 of \cite{Plakias:2023esq}. As expected, the strongest constraint comes from $\text{Br}( K_L \rightarrow e \mu) < 6.3 \times 10^{-12}$ \cite{ParticleDataGroup:2022pth}, which plateaus for $m_a \lesssim m_{K_L}$, setting $\vphi \gtrsim 10$ TeV. Bounds from other scalar meson decays are at least two orders of magnitude less stringent, while vector meson decays are negligible due to their large width. For completeness, our numerical analysis includes constraints from unflavored meson decays. The expressions differ slightly due to the anomaly contribution~\cite{Cornella:2019uxs}. However, the large width of these mesons leads to much weaker bounds.

Lepton flavor conserving decays also play an important role, as the diagonal couplings in Eq.~\eqref{eq:s3:cecuMat} are larger than the off-diagonal ones, which are further suppressed by the small hierarchical mixing. The expression for the total decay rate is complicated by the interference between the SM and ALP contributions. To handle this properly, we compute the corresponding bounds using \texttt{flavio}~\cite{Straub:2018kue}, leveraging the fact that in these processes, the virtual momentum of the ALP is essentially fixed by the meson mass. Our findings show that the strongest constraint arises from $K_L \rightarrow \mu\mu$, which plateaus at roughly $\vphi \gtrsim 70$\,TeV.

Semileptonic decays are enhanced in the on-shell ALP production regime $m_{M_1} -m_{M_2}> m_a > m_{\ell_i} + m_{\ell_j}$, in which the branching ratio can be factorized as 
\begin{align}
    \text{Br}( M_1\rightarrow M_2 ^{(*)} \, \ell_i \bar \ell_j ) \simeq \text{Br}(M_1\rightarrow M_2 ^{(*)} \, a) \text{Br}(a\rightarrow  \ell_i \bar \ell_j ),
\label{eq:s3:semileptonicFormula}
\end{align}
where
\begin{align}
    \begin{aligned}
    \Gamma (a\rightarrow f_i \bar f_j) = m_a \frac{N_c}{16\pi } &\lambda^{1/2}\left(\frac{m_i}{m_a},\frac{m_j}{m_a}\right) \\
    & \times \left[\left(|c_{ij}^f|^2+|c_{ji} ^f|^2\right)\left(1- \frac{m_i^2+m_j^2}{m_a^2}  \right) + 4 \Re c_{ij} ^f c_{ji} ^{f} \frac{m_i m_j}{m_a^2} \right].
    \end{aligned}
    \label{eq:s3:Gammaaff}
\end{align}
Defining $\braket{M_2 ^{*} (p_2) | \bar q_k \gamma^5 q_l | M_1 (p_1)} = -2 i  f^* (q^2) _{M_1\rightarrow M_2 ^*} (\eps \cdot q) m_{M_2 ^*}/(m_{q_k} + m_{q_l})$ with $q=p_1-p_2$ and $\eps_\mu=\eps_\mu(p_2)  $ the polarization vector of $M_2 ^*$, one finds
\begin{align}
    \begin{aligned}
    \Gamma (M_1 \rightarrow M_2 ^* a) = \frac{m_{M_1}^3}{64\pi }\lambda^{3/2}\left(\frac{m_{M_2 ^*}}{m_{M_1}}, \frac{m_{a}}{m_{M_1}} \right)& \frac{|f^* _{M_1\rightarrow M_2 ^*} (m_a^2)| ^2}{(m_{q_k}+m_{q_l})^2} \\
    & \times  \left[|c_{kl} ^f|^2 +|c_{lk} ^f|^2  +2 \Re c_{kl} ^f c_{lk} ^f \right].
    \end{aligned}
\end{align}
In this category, we can identify two classes of experimental searches. Searches for prompt decays reconstruct the 3-body vertex with a typical resolution of approximately $\mathcal{O}(100) \, \mu$m. When applying constraints from such searches, we require $ l_a \lesssim 100 \, \mu\text{m} $, unless stated otherwise. Searches for displaced vertices, instead, provide bounds based on the lifetime of the long-lived particle. We will distinguish between these two types of searches and their corresponding bounds in the following discussion.

Let us first consider \textit{prompt} decays. For lepton flavor-violating ones, we revisit the processes studied in Ref.~\cite{Plakias:2023esq}. The strongest constraint comes from $\text{Br}(B \rightarrow K\mu \tau) < 7.7 \times 10^{-6}$ \cite{Belle:2022pcr}, followed by $\text{Br}(B\rightarrow \pi \mu \tau)<5.9\times 10^{-5}$ \cite{BaBar:2012azg}, and $\text{Br}(B \rightarrow K^{*} \mu \tau)<9.8 \times 10^{-6}$ \cite{LHCb:2022wrs}.
For lepton flavor conserving decays, the situation is complicated by current tensions between SM predictions and experimental values of various branching ratios and angular observables involving the underlying $b \to s \mu \mu$ transition~\cite{Greljo:2022jac, Alguero:2023jeh, Ciuchini:2022wbq, Hurth:2023jwr, Guadagnoli:2023ddc}. In contrast, the theoretically cleaner $R_{K^{(*)}}$ ratios are measured to be consistent with SM expectations~\cite{LHCb:2022qnv, LHCb:2022vje}. Here, we only consider these ratios to derive constraints on the ALP contributions, roughly setting $\vphi \gtrsim$~TeV.

For \textit{displaced} vertex signatures, we consider the bounds from LHCb on $B\rightarrow K^{(*)}\mu\mu$ \cite{LHCb:2015nkv, LHCb:2016awg}, which sets exclusion limits in the region $10^3 \text{ GeV} \lesssim \vphi \lesssim 10^7$ GeV, see also Ref.~\cite{Gavela:2019wzg}. 
Additionally, we consider the decays $B\rightarrow K\gamma\gamma$ and $K\rightarrow \pi \gamma\gamma$, computed analogously using the well-known expressions for $\text{Br}(a\rightarrow\gamma \gamma)$ found in \cite{Bauer:2017ris}.
We employ the bound on $\text{Br}(K\rightarrow \pi \gamma \gamma)$ for long-lived ALPs from NA62 \cite{NA62:2023olg} and the bound on $\text{Br}(B\rightarrow K \gamma \gamma)$ from BaBar \cite{BaBar:2021ich}. These constraints are similar, setting $\vphi \gtrsim$\,TeV, with the former being slightly stronger but applicable over a much narrower interval of ALP masses.

\noindent
\emph{VLF}. Let us now turn to the UV contributions obtained by integrating out VLFs to generate local operators that directly affect rare meson decays. As mentioned earlier, the leading effects come from operators not involving the $\Phi$ field. In particular, the Higgs-current operators
\begin{equation}
\label{eq:s3:OHdOHu}
\begin{split}
    \mathcal{O}_{H d}^{ij} &= (H^\dagger i \overset{\leftrightarrow}{D}_\mu H) (\bar{d}_R^i \gamma^\mu d_R^j)\,, \\
    \mathcal{O}_{H u}^{ij} &= (H^\dagger i \overset{\leftrightarrow}{D}_\mu H) (\bar{u}_R^i \gamma^\mu u_R^j)\,,
\end{split}
\end{equation}
are generated already at the tree level integrating out the heavy $Q_2^a$. The corresponding Wilson coefficients read
\begin{equation}
\label{eq:VLQ_CHdCHu}
    C_{H d}^{ij} = \frac{y_j^{da} y_{i}^{da\ast}}{2M_{Q_2}^2}\,, \qquad 
    C_{H u}^{ij} = \frac{y_j^{ua} y_{i}^{ua\ast}}{2M_{Q_2}^2} \, .
\end{equation}
These operators induce flavor-changing $Z$-boson couplings to right-handed quarks after electroweak symmetry breaking~\cite{Jenkins:2017jig}. Focusing on down quark transitions, we define the weak effective Hamiltonian for $d_j\to d_i \ell \ell$ and $d_j\to d_i \nu \nu$ as $\mathcal{H}_\mathrm{eff}= - \mathcal{N} \sum_i C_i O_i + \text{h.c.}$,
where $\mathcal{N} = \frac{4 G_F}{\sqrt{2}} \frac{e^2}{16\pi^2} V_{tj} V_{ti}^\ast$ and the sum runs over the semileptonic local operators. Then, the $\mathcal{O}_{H d}$ operator is matched by a $Z$ exchange to
\begin{equation}
C_{10}^{\prime ij \ell\ell} = \frac{1}{2\mathcal{N}} C_{Hd}^{ij}\,, \qquad C_R^{ij\nu\nu} = \frac{1}{\mathcal{N}} C_{Hd}^{ij}\,,
\end{equation}
with $O_{10}^{\prime ij\ell\ell} =  (\bar{d}_i \gamma_\mu P_R d_j) (\bar{\ell} \gamma^\mu \gamma_5\ell)$\footnote{The related operator involving a vector leptonic current $O_{9}^{\prime}$ is suppressed due to the small SM $Z$ couplings to these leptonic currents.} and $O_{R}^{ij\nu\nu} = (\bar{d}_i \gamma_\mu P_R d_j) (\bar{\nu}\gamma_\mu (1-\gamma_5) \nu)$. We stress that these contributions are lepton flavor universal. We use \texttt{flavio}~\cite{Straub:2018kue} to obtain the bounds from meson decays with charged or neutral leptons in the final states. The leading constraints are due to kaon decays, in particular to $K\to \pi \nu \nu$ setting the leading constraint $\vphi \gtrsim 200~\mathrm{GeV}$. Purely leptonic kaon decays $K_L\to ee, \mu\mu$ offer a slightly worse sensitivity, $\vphi \gtrsim 100~\mathrm{GeV}$. $B$ and $D$ decays lead to subleading constraints.

Lepton flavor-violating contributions can arise in two ways. The first is via the exchange of a $Z$ boson, with one insertion of Eq.~\eqref{eq:s3:OHdOHu} and one of
\begin{align}
\begin{aligned}
    \mathcal{O}_{H \ell}^{(1)ij} &= (H^\dagger i \overset{\leftrightarrow}{D}_\mu H) (\bar{\ell}_R^i \gamma^\mu \ell_R^j) \\
    \mathcal{O}_{H \ell}^{(3)ij} &= (H^\dagger i \overset{\leftrightarrow}{D}_\mu^I H) (\bar{\ell}_R^i \tau^I \gamma^\mu \ell_R^j)\,,
    \label{eq:s3:OHl}
\end{aligned} 
\end{align}
generated at tree level via the heavy $E_2 ^a$. The corresponding Wilson coefficients are
\begin{equation}
    C_{H \ell}^{(1)ij} = C_{H \ell}^{(3)ij} = -\frac{y_i^{ea} y_{j}^{ea\ast}}{4M_E^2}\,.
    \label{eq:s3:Hlcoeff}
\end{equation}
The second way involves integrating out the VLFs at 1-loop, directly generating the 4-fermion SMEFT operator
\begin{align}
    \mathcal{O}^{ijkl} _{\ell d} = (\bar \ell_i \gamma ^\mu \ell_j) (\bar d_k \gamma _\mu d_l)\,,
    \label{eq:s3:Olldd}
\end{align}
with Wilson coefficient\footnote{We used Matchete~\cite{Fuentes-Martin:2022jrf} to obtain this and the other 1-loop Wilson coefficients presented in the paper, cross-checking the expressions with known results in the literature.}
\begin{align}
    \begin{aligned}
        C^{ijkl} _{\ell d} =\frac{1}{128\pi^2} &\left[2
        y_{i}^{ea}  (y_{j}^{ea})^* (y_{k}^{da})^*  y_{l}^{da}   \frac{\log M_{E_2}^2/M_{Q_2}^2}{M_{E_2}^2 - M_{Q_2}^2}  \right. \\
        &  \left. - \frac{3}{M_{E_2}^2} y_{i}^{ea}  (y_{j}^{ea})^* (z_{k}^{d})^* z_{l}^{d}  
         - \frac{3}{M_{Q_2}^2}
        z_{i}^{ea}  (z_{j}^{ea})^* (y_{k}^{da})^*  y_{l}^{da} 
        \right]\,.
    \end{aligned}
    \label{eq:s3:Clldd}
\end{align}
Despite the 1-loop factor, this provides the dominant constraint since $M_{Q_2,E_2} \gtrsim 4 \pi v$. In principle, there is a third way via SMEFT (renormalization group) RG effects on Eq.~\eqref{eq:s3:OHl}. However, the RG effect is always subleading, as flavor mixing would be CKM-suppressed and thus less significant compared to the anarchic nature of the coefficient in Eq.~\eqref{eq:s3:Clldd}.
To establish the bounds, we map the contribution of these Wilson coefficients to the same set of lepton flavor violating decays analyzed in the ALP-mediated case, following~\cite{Plakias:2023esq}. The resulting constraints are significantly weaker compared to the lepton flavor-conserving ones obtained earlier, with the strongest being associated with $K_L \rightarrow e\mu$, leading to $\vphi \gtrsim 10$\,GeV.

\subsubsection*{Neutral meson mixing}

One of the leading constraints for both flavorful ALP and VLQs is due to their contributions to the CP-even and CP-odd observables measured in $P^0-\bar{P}^0$ oscillations, with $P=K, B, B_s, D$. We use \texttt{flavio} to obtain constraints on both the ALP and the VLQ contributions to the following observables: $\epsilon_K$, measuring CP violation in kaon mixing~\cite{Buras:2010pza, Blum:2011ng, Brod:2011ty, Garron:2016mva, ParticleDataGroup:2022pth, FlavourLatticeAveragingGroupFLAG:2021npn},\footnote{Note that we do not use $\Delta m_K$ as a constraint due to the large uncertainties in its SM prediction from long-distance contributions. Instead, we use its experimental value as an input when predicting $\epsilon_K$. Nevertheless, we checked that saturating the experimental value of $\Delta m_K$ with the NP contributions leads at most to bounds comparable to that from $\epsilon_K$.} $\Delta m_s$ ($\Delta m_d$) and $S_{\psi \phi}$ ($S_{\psi K}$) measuring the neutral meson mass difference and mixing induced CP violation in $B_s$ ($B_d$) mixing~\cite{King:2019lal, Dowdall:2019bea, Greljo:2022jac, HFLAV:2022esi}, and $x_{12}^{\mathrm{Im}, D}$ measuring CP violation in $D$ mixing~\cite{Carrasco:2015pra, HFLAV:2022esi}.

\noindent
\emph{ALP}. We write the $\Delta F=2$ Hamiltonian in the quark mass basis as
\begin{align}
\begin{aligned}
\mathcal{H}^{\Delta F=2}_\mathrm{NP} &= C_1 ^{ij} (\bar q_L ^i \gamma ^\mu q_L ^j)^2 + \widetilde C_1 ^{ij} (\bar q_R ^i \gamma ^\mu q_R ^j)^2   + C_2 ^{ij} (\bar q_R ^i q_L ^j)^2 + \widetilde C_2 ^{ij} (\bar q_L ^i q_R ^j)^2\\
&+ C_4 ^{ij} (\bar q_R ^i q_L ^j) (\bar q_L ^i q_R ^j) + C_5 ^{ij} (\bar q_L ^i \gamma ^\mu q_L ^j) (\bar q_R ^i \gamma _\mu q_R ^j) +\text{h.c.}\,,
\end{aligned}
\end{align}
with $q=u, d$ and $i,j$ are flavor indices, whereas the color indices are contracted inside parentheses. Integrating out the scalar and pseudoscalar components of $\Phi$ generates $\Delta F=2$ operators already at the tree level as follows
\begin{align}
\begin{aligned}
    C_2 ^{q,ij} &= - \frac{(c_{ji}^{q*})^2}{2} \left( \frac{1}{m_\rho ^2} - \frac{1}{m_a^2} \right)\,, \\
    \widetilde C_2 ^{q,ij} &= - \frac{(c_{ij} ^q )^2}{2} \left( \frac{1}{m_\rho ^2} - \frac{1}{m_a^2} \right)\,, \\
    C_4 ^{q,ij} &= - c_{ij} ^q c_{ji}^{q *}\left( \frac{1}{m_\rho ^2} + \frac{1}{m_a^2} \right)\,. \\
\end{aligned}
\end{align}
Here we have assumed the ALP to be heavier than the appropriate scale of each $P-\bar{P}$ process. In the case of a light ALP, this is no longer applicable, and ALP propagator effects should be included~\cite{Bauer:2021mvw}. However, in this work, we do not consider the meson mixing constraints in the case of a light ALP, as we find that parameter space is predominantly constrained by other processes, such as rare meson decays. Outside of that region, we anticipate that meson mixing observables are responsible for the strongest bound coming from ALP physics. This roughly sets $\vphi \gtrsim 10 \text{ TeV} \times (10 \text{ GeV}/m_a)$.

\noindent
\emph{VLF}.
Among the VLQs with the interaction Lagrangian defined in Eq.~\eqref{eq:s3:L_VLFs}, $Q_2^a$ can generate $\Delta F=2$ operators via two effects. The first effect is obtained by integrating out $Q_2^a$ at the tree level, as discussed in the rare meson decay case; see Eq.~\eqref{eq:VLQ_CHdCHu}. The generated operators could, in principle, generate $\Delta F=2$ operators via two insertions of the modified $Z$ couplings. Such contributions are highly suppressed. Important effects can, however, be generated by 1-loop SMEFT RG. Performing one insertion of \eqref{eq:VLQ_CHdCHu} and closing the loop with two Higgses and a fermion, we obtain $y_t$-enhanced effects by generating the left-right 4-fermion $\mathcal{O}_{qd}$ operator  (in the notation of \cite{Grzadkowski:2010es}). These effects have been found to be phenomenologically relevant in Ref.~\cite{Bobeth:2016llm}, and we take them into account in our analysis.

The second way to generate $\Delta F=2$ effects is from one-loop matching contributions. This directly generates 4-fermion operators required to induce such effects:
\begin{align}
    \begin{aligned}
    \mathcal{O}_{dd} ^{ijkl}&=(\bar d_i \gamma ^\mu d_j) (\bar d_k \gamma _\mu d_l),\\
    \mathcal{O}_{uu}^{ijkl}&=(\bar u_i \gamma ^\mu u_j) (\bar u_k \gamma _\mu u_l).
    \end{aligned}
    \label{eq:s3:Odddd&Ouuuu}
\end{align}
The Wilson coefficients associated with these operators read
\begin{align}
    \begin{aligned}
    C^{ijkl}_{dd}&=\frac{1}{64\pi^2} \frac{1}{M_{Q_2}^2} (y_{i}^{da})^* y_{j}^{da}\left[ -(y_{k}^{db})^* y_{l}^{db}  + 3 (z_s ^d)^* z_t ^d \right],\\
    C^{ijkl}_{uu}&=\frac{1}{64\pi^2} \frac{1}{M_{Q_2}^2} (y_{i}^{ua})^* y_{j}^{ua}\left[ -(y_{k}^{ub})^* y_{l}^{ub}  + 3 (z_s ^u)^* z_t ^u \right].
    \end{aligned}
\end{align}
The $\Delta F=2$ operators are obtained by selecting $i=k$ and $j=l$. In the minimal basis defined in App.~\ref{app:paramsMatching}, the terms involving $z$ couplings do not contribute. 

Employing the benchmark discussed in App.~\ref{app:paramsMatching}, we find that these one-loop matching contributions offer the dominant constraint, which comes from $\epsilon_K$, setting $v_\Phi \gtrsim 2.8~\mathrm{TeV}$. The rest of the $\Delta F=2$ constraints, namely the CP-even and CP-odd observables in $B_d$, $B_s$ and $D$-mixing, show various degrees of interplay between tree-level and one-loop contributions of VLQs, but offer only subleading constrains of $v_\Phi \gtrsim 100~\mathrm{GeV}$.

It is worth mentioning that also the operators $\mathcal{O}_{qd},\mathcal{O}_{qu}$ can be generated at the one-loop level and could, in principle, lead to important effects. Indeed, the components diagonal in $q$ do not need an insertion of $\Phi$ and hence can be generated already at dimension 6. After electroweak symmetry breaking, the rotation to the mass basis induces additional flavor violation in the down-strange sector, suppressed only by the CKM. This suppression could, in principle, be overcome by an enhancement in the observables due to the different chirality structures of the operators in such a way that the bounds become potentially important. 
Accidentally, in $\Z_4$ models the operator $(\bar q_2 \gamma ^\mu q_2)(\bar d_i \gamma_\mu d_j) $ is generated at 1-loop with coefficient $\sim (y^{da}_i x_{2} ^{qa})^* (y^{da}_j x_{2} ^{qa})$ (and similarly for up-quarks), which vanishes whenever the first generator is involved (see App.~\ref{app:paramsMatching}). Hence, this effect could only appear in $B_s$-mixing, which is not very constraining.

\subsubsection*{$\bm{\epsilon^\prime/\epsilon}$}
The $\epsilon^\prime/\epsilon$ observable acts as a probe of direct CP violation in $K_L \to \pi \pi$. The current best estimate of the SM predictions is $(\epsilon^\prime/\epsilon)_\mathrm{SM} = (13.9\pm5.2)\times 10^{-4}$~\cite{RBC:2020kdj, Buras:2020pjp, Aebischer:2020jto, Aebischer:2021hws, Cirigliano:2019cpi}, while the experimental world average is $(\epsilon^\prime/\epsilon)_\mathrm{exp} = (16.6\pm2.3)\times 10^{-4}$~\cite{ParticleDataGroup:2022pth}. We use the master formula presented in Refs.~\cite{Aebischer:2021hws,Aebischer:2018quc} to compute the BSM contributions to $\epsilon^\prime/\epsilon$ as
\begin{equation}
    \left(\frac{\epsilon^\prime}{\epsilon} \right)_\mathrm{BSM} = \sum_b P_b (\mu_\mathrm{ew}) \mathrm{Im} \left[C_b(\mu_\mathrm{ew}) - C_b^\prime(\mu_\mathrm{ew}) \right]\,,
\end{equation}
where $P_b$ parameterizes the matrix elements of the relevant four quark operators, which we consider in the JMS basis~\cite{Jenkins:2017jig}. We use the numerical values reported in Table 3 of Ref.~\cite{Aebischer:2021hws}.

\noindent
\emph{ALP}. The ALP contributes to $\epsilon'/\epsilon$ at the tree-level through scalar four-fermion $\Delta F=1$ operators. In the notation of Ref.~\cite{Aebischer:2021hws} these are $[O_{dd}^{S1,RR}]_{2111}$, $[O_{ud}^{S1,RR}]_{1121}$, and their $L\leftrightarrow R$ counterparts. However, its contributions are suppressed by first-generation quark masses. At most, this renders only subleading constraint up to $\vphi \gtrsim 100 \text{ GeV}$.

\noindent
\emph{VLF}. VLQs induce flavor-violating $Z$-boson couplings to right-handed quarks via Eq.~\eqref{eq:VLQ_CHdCHu}. As can be seen from Table~$3$ of Ref.~\cite{Aebischer:2021hws}, the largest matrix elements are due to left-right vector operators. In our setup, $[O_{dd}^{V1,LR}]_{2111}$, $[O_{du}^{V1,LR}]_{2111}$ are generated by integrating out the $Z$ boson, with VLQ-induced flavor violating right-handed down quark vector currents, contracted with either left-handed up or down quark currents due to the SM $Z$ couplings. We obtain the bound $\vphi \gtrsim 0.7~\text{TeV}$.

\subsubsection*{$\bm{\ell_i \to \ell_j \gamma}$}

Both the ALP and the VLFs generate lepton flavor-violating couplings that are strongly constrained by purely leptonic observables. Dipole transitions can be encoded in the WET Lagrangian via 
\begin{equation}
    \mathcal{L}_D \supset C_{e\gamma}^{ij} \left(\bar{e}_i \sigma^{\mu\nu} P_L e_j\right) F^{\mu\nu} + \mathrm{h.c.}\,.
    \label{eq:s3:WETdipole}
\end{equation}
Assuming $m_{\ell_j}>m_{\ell_i}$, the branching ratio of $\ell_j \to \ell_i \gamma$ then reads~\cite{Lavoura:2003xp}
\begin{equation}
    \mathrm{Br}(\ell_j \to \ell_i \gamma) = \frac{m_{\ell_i}^3}{4\pi \Gamma_{\ell_j}} \left(|C_{e\gamma}^{ij}|^2 + |C_{e\gamma}^{ji}|^2 \right)\,.
\end{equation}
\emph{ALP}. The ALP $a$ generates these observables at the one loop. In the regime in which the ALP is much more massive than $\ell_j$, the Wilson coefficient associated with the operator \eqref{eq:s3:WETdipole} reads
\begin{equation}
\begin{split}
    C_{e\gamma}^{ij} &= \frac{e}{32\pi^2 m_a^2} \sum_{k=1,2,3} \left[ \frac{1}{6} \left(m_{\ell_j} c_{ki}^{e\ast} c_{kj}^{e} + m_{\ell_i}  c_{ik}^{e} c_{jk}^{e\ast}\right)-
    m_{\ell_k}  c_{ki}^{e\ast} c_{jk}^{e\ast}\left(\frac{3}{2}+\log{\frac{m_{\ell_j} ^2}{m_a^2}}\right) 
    \right] \,
\end{split}
\label{eq:s3:CegammaALP}
\end{equation}
whereas the contribution of $\rho$ can be obtained by the replacement $m_a^2 \to m_\rho^2$, $c\to -i c$. The contribution from Barr-Zee diagrams is relevant only for ALP masses below the charm mass, given the smallness of our ALP couplings with the top and bottom quarks. The expression for these contributions can be found in \cite{Bauer:2021mvw}, and we take them into account in our final plot.
The relevant current experimental bounds are $\text{Br} (\mu \rightarrow e \gamma) < 4.2 \times 10^{-13}$~\cite{MEG:2013oxv,MEG:2016leq} and $\text{Br} (\tau \rightarrow \mu \gamma) < 4.4 \times 10^{-8}$~\cite{BaBar:2009hkt, Belle:2007qih}. Both lead to comparable bounds due to the hierarchical structure of $c^e _{ij}$ up to $\vphi \gtrsim$ TeV.

\noindent
\emph{VLF}. The operator \eqref{eq:s3:WETdipole} can also be generated by integrating out the heavy VLFs at 1-loop. The corresponding $\Z_4$-invariant SM$+\Phi$ terms are
\begin{align}
    C_{eB\Phi} ^{ij} \, \Phi^{n^e_{ij}}\bar \ell_i \sigma^{\mu\nu} H e_{j} B_{\mu\nu} + 
    C_{eW\Phi} ^{ij} \, \Phi^{n^e_{ij}}\bar \ell_i \sigma^{\mu\nu} \tau^a H e_{j} W_{\mu\nu} ^a\,,
    \label{eq:s3:OeBPhi}
\end{align}
where by simple dimensional analysis $C_{eB\Phi,eW\Phi} ^{ij} \sim  g^{(\prime)}  y_* ^{3+ n^e_{ij}}/ 16\pi^2 M ^{2+n^e _{ij}}$, with $y_*$ some representative UV coupling. After $\Phi$ condensation, this leads to the SMEFT operators $\mathcal{O}_{eB, eW}$,
which combine to Eq.~\eqref{eq:s3:WETdipole}. Using Matchete \cite{Fuentes-Martin:2022jrf}, we were able to exactly extract the Wilson coefficients relevant for $\mu \rightarrow e\gamma$ and $\tau \rightarrow \mu \gamma$:
\begin{align}
\begin{aligned}
   C_{eB\Phi} ^{i2} &= \frac{g'}{32\pi^2} \frac{1}{M_{E_2}^2} \left[-\frac{1}{M_{E_1}} (x^{ea} _{12})^* x_2 ^{ea}   x_{12}^{eb} y^{eb}_i \right. \\
   &  \qquad \qquad \qquad \left. + \frac{1}{12} \frac{1}{M_{E_2}}  \left( x_2 ^{ea}  (x_2 ^{ea})^*x_2 ^{eb} y_i ^{eb} + 2 x_2 ^{ea}  y_{k} ^{ea}  (y_k ^{eb})^*  y_i ^{eb} \right)  \right], \\
   C_{eW\Phi} ^{i2} &= \frac{g}{384 \pi^2} \frac{1}{M_{E_2}^3 }   x_2 ^{ea}  y_{k} ^{ea}  (y_k ^{eb})^*  y_i ^{eb}\,.
\end{aligned}
\label{eq:s3:CeBWPhiMatchete}
\end{align}
The coefficients $C_{eB\Phi, eW\Phi}^{i3}$ are null because $e_3$ does not couple to any VLF, while $C_{eB\Phi, eW\Phi}^{i1}$ arise at dimension 8 since they involve a further power of $\Phi$. Hence, $ C_{eB\Phi, eW\Phi}^{i2}$ provide the leading contributions to the aforementioned transitions. Note that the couplings in Eq.~\eqref{eq:s3:CeBWPhiMatchete} are not perfectly aligned with the effective $Y_e$ (see App.~\ref{app:paramsMatching}). This leads to a further mixing between the components of the dipoles after the diagonalization of the Yukawa with a $e_i$ rotation.  We match these coefficient to $C_{e\gamma}$ after $\Phi$ condensation via $C_{e\gamma} ^{ij} = v(c_w C_{eB} - s_w C_{eW})/\sqrt{2}$. The coefficient qualitatively scales as $C_{e\gamma}^{i2} \sim e v y_* ^4 \eps ^3 / 16\pi^2 \vphi^2$, that can lead to bounds on $\vphi$ of $\mathcal{O}(\text{TeV})$ for $y_* \sim \Oone$ due to the stringent constraint on $\mu \rightarrow e \gamma$. Precisely, employing the benchmark in App.~\ref{app:paramsMatching}, we find $\vphi \gtrsim 150$\,GeV.

\subsubsection*{$\bm{\mu \to e}$ conversion in nuclei}
$\mu\rightarrow e$ conversion in nuclei is one of the most constraining observables on new physics models predicting lepton flavor violation. Following Ref.~\cite{Crivellin:2017rmk}, the operators relevant to our discussion comprise the following set
\begin{equation}
\label{eq:mutoeOps}
\begin{split}
    O_{qq}^{VLL} &= (\bar{e} \gamma_\mu P_L \mu) (\bar{q} \gamma^\mu P_L q) \,, \\
    O_{qq}^{VLR} &= (\bar{e} \gamma_\mu P_L \mu) (\bar{q} \gamma^\mu P_R q) \,, \\
    O_{qq}^{SLL} &= (\bar{e} P_L \mu) (\bar{q} P_L q) \,, \\
    O_{qq}^{SLR} &= (\bar{e} P_L \mu) (\bar{q} P_R q) \,, \\
    O_{gg}^{L} &= \alpha_s m_\mu G_F (\bar{e} P_L \mu) G_{\mu\nu}^a G^{a \mu\nu} \,,
\end{split}
\end{equation}
together with their $L\leftrightarrow R$ equivalents and the dipole operators of Eq.~\eqref{eq:s3:WETdipole}.\footnote{As for $\ell_i\rightarrow \ell_j \gamma$ processes, the contributions from Barr-Zee diagrams are relevant only for $m_a\lesssim m_c$. We take this into account in our final plot, taking the expression for these contributions from \cite{Bauer:2021mvw}.} The conversion rate on a nucleus $N$ is \cite{Kitano:2002mt, Cirigliano:2009bz, Crivellin:2017rmk, Shifman:1978zn}
\begin{equation}
\label{eq:Gammamutoe}
    \Gamma_{\mu \to e}^N = \frac{m_\mu^5}{4} \left|
    C_L^D D_N + 2 \left(G_F m_\mu m_p \tilde{C}_{(p)}^{SL} S_N^{(p)}
    +\tilde{C}_{(p)}^{VR} V_N^{(p)} + p\to n\right)
    \right|^2 + L \leftrightarrow R\,,
\end{equation}
where $C_L^D \equiv C_{e\gamma}^{e\mu}/m_\mu$ and
\begin{equation}
\begin{split}
    \tilde{C}_{p/n}^{VR} &= \sum_{q=u,d} \left(C_{qq}^{V RL} + C_{qq}^{V RR} \right) f_{V p/n}^{(q)} \,, \\
    \tilde{C}_{(p/n)}^{SL} &= \sum_{q=u,d,s} \frac{\left(C_{qq}^{S LL} + C_{qq}^{S LR}\right)}{m_\mu m_q G_F} f_{Sp/n}^{(q)} + \tilde{C}_{gg}^{L} f_{Gp/n} \,, \\
    \tilde{C}_{gg}^{L} &= C_{gg}^{L} - \frac{1}{12 \pi} \sum_{q=c,b} \frac{\left(C_{qq}^{S LL} + C_{qq}^{S LR}\right)}{m_\mu m_q G_F} \,,
\end{split}
\end{equation}
and similarly for $L\leftrightarrow R$. The vector operator nucleon form factors are $f_{Vp}^{(u)}=2$, $f_{Vn}^{(u)}=1$, $f_{Vp}^{(d)}=1$, $f_{Vn}^{(d)}=2$, while for the scalar and the gluonic form factors, we take the values as reported in Ref.~\cite{Crivellin:2017rmk}, see also Refs.~\cite{Crivellin:2014cta, Hoferichter:2015dsa, Crivellin:2013ipa,Junnarkar:2013ac}. The most sensitive current measurement is on $\mathrm{Au}$ for which we have the overlap integrals $V_{\mathrm{Au}}^{(p)} = 0.0974$, $V_{\mathrm{Au}}^{(n)} = 0.146$, $S_{\mathrm{Au}}^{(p)} = 0.0614$, $S_\mathrm{Au}^{(n)} = 0.0918$, $D_\mathrm{Au}=0.189$ ~\cite{Kitano:2002mt}.
Finally, the prediction should be compared to the 90\% CL limit on $\mu \to e$ conversion in gold by SINDRUM II~\cite{SINDRUMII:2006dvw}
\begin{equation}
    \mathrm{Br} (\mu \to e) = \frac{\Gamma(\mu^- \mathrm{Au} \to e^- \mathrm{Au})}{\Gamma_\mathrm{capt}(\mu^- \mathrm{Au})} < 7 \times 10^{-13}\,,
\end{equation}
where the estimate for the capture rate can be found in~\cite{Suzuki:1987jf, Kitano:2002mt}. 

\noindent
\emph{ALP}. The ALP contributes to $\mu \to e$ conversion in multiple ways: directly generating the scalar four fermion operators $O_{qq}^{S}$ at the tree level, contributing through the dipole operator as in Eq.~\eqref{eq:s3:CegammaALP}, or generating $O_{gg}$ after integrating out the heavy SM quarks. The tree-level matching results in 
\begin{equation}
\begin{split}
    C_{qq}^{SLL} &= c^{q \ast}_{qq} c^{e \ast}_{21} \left(\frac{1}{m_\rho^2} - \frac{1}{m_a^2}\right) \,,\\
    C_{qq}^{SLR} &= c^{q}_{qq} c^{e \ast}_{21} \left(\frac{1}{m_\rho^2} + \frac{1}{m_a^2}\right)\,,\\
    C_{qq}^{SRL} &= c^{q \ast}_{qq} c^{e}_{12} \left(\frac{1}{m_\rho^2} + \frac{1}{m_a^2}\right)\,,\\
    C_{qq}^{SRR} &= c^{q}_{qq} c^{e}_{12} \left(\frac{1}{m_\rho^2} - \frac{1}{m_a^2}\right)\,,\\
\end{split}
\end{equation}
Note that for coherent conversion in Eq.~\eqref{eq:Gammamutoe} only the scalar quark current operator enters, thus the pseudoscalar $a$ contributions above cancel as $c_{qq}^q$ are real. We obtain bounds of up to $\vphi\gtrsim$ TeV.

\noindent
\emph{VLF}. Muon conversion into an electron in the field of a nucleus can also proceed through SM $Z$ exchange, with flavor-violating couplings to leptons induced by the operator in Eq.~\eqref{eq:s3:OHl} and exploiting the SM coupling to quarks. The vector operators in Eq.~\eqref{eq:mutoeOps} are generated with~\cite{Crivellin:2020ebi}
\begin{align}
\begin{aligned}
    C_{qq}^{VLL} &= \Gamma_{e\mu}^{\ell L} \frac{1}{M_Z^2} \Gamma_{qq}^{L} \,, \\
    C_{qq}^{VLR} &= \Gamma_{e\mu}^{\ell L } \frac{1}{M_Z^2} \Gamma_{qq}^{R} \,.
\end{aligned}
\end{align}
The $Z$ boson couples to quarks via $\Gamma_{qq}^{L,R}$ in the standard way, whereas the $Z\mu e$ coupling reads
\begin{equation}
\label{eq:S3:VLLcLFV}
    \Gamma_{e\mu}^{\ell L} = \frac{g_2}{2 c_W} v^2 \left(C_{H\ell}^{(1)12} +C_{H\ell}^{(3)12} \right)\,.
\end{equation}
Using our benchmark, we obtain the bound $\vphi \gtrsim 1.1~\mathrm{TeV}$.

\subsubsection*{Rare lepton decays}
Rare lepton decays comprise a wide set of observables, including flavor-violating $\ell_i \rightarrow \ell_j +\text{invisible}$, $\ell_i \rightarrow \ell_j \ell_k\ell_k$ and $\ell_i \rightarrow M^{(*)} \ell_j$. As usual, we will focus on ALP and VLF contributions separately.

\noindent
\emph{ALP}.
As with mesons, we begin by discussing flavor-violating decays with an invisible ALP as a final state. Bounds on the decay of a muon into an electron and an invisible scalar have been set at TRIUMF \cite{Jodidio:1986mz} and from TWIST \cite{TWIST:2014ymv}, while analogous bounds for $\tau$ decays have been reported by Belle II \cite{Belle-II:2022heu}. The relevant formula for such decays is
\begin{align}
    \Gamma(f_i \rightarrow f_j a) = m_i \frac{N_c}{32\pi} \lambda^{1/2}\left(\frac{m_a}{m_i},\frac{m_j}{m_i}\right) \left[\left(|c_{ij}^f |^2 + |c_{ji} ^f|^2\right)  \left( 1-\frac{m_a^2-m_j^2}{m_i^2}\right)  - 4 \Re c_{ij}^f c_{ji}^f \frac{m_j}{m_i}  \right],
    \label{eq:s3:Gammafaf}
\end{align}
and we again impose $\ell_a \gtrsim 1$\,m for the ALP to be considered invisible. The bounds on $\vphi$ that we obtain are almost always superseded by the one coming from $K\rightarrow \pi a$.

Next we consider the $\ell_i \to M^{(*)} \ell_j$ processes, where $M$ is a scalar (vector) meson. In the ALP-mediated case, the decay rate for decays in flavored scalar mesons closely resembles one of the leptonic flavored meson decays \eqref{eq:s3:leptonicMesonDecaysRate}:
\begin{align}
\begin{aligned}
    \Gamma\left(\ell_i \rightarrow M \ell _j \right) = \frac{f_M ^2 m_{\ell_i} }{128\pi }& \frac{\lambda^{1/2} \left( \frac{m_M}{m_{\ell_i}},  \frac{m_{\ell_j}}{m_{\ell_i}} \right)}{(m_{q_k} + m_{q_l})^2} \frac{\left[ |c^q_{kl}|^2 + |c^q_{lk}| +2 \Re c^q_{kl} c^q_{lk}\right]}{\left(1-m_{a}^2/m_M^2 \right)^2 + m_a ^2 \Gamma_a ^2/m_M^4 } \\
    & \times \left[ \left( |c^e_{ij}|^2 + |c^e_{ji}|^2\right) \left( 1- \frac{m_M ^2 - m_{\ell_j}^2}{m_{\ell_i}^2} \right) -4 \Re c^e _{ij} c^e _{ji} \frac{ m_{\ell_j} }{m_{\ell_i}} \right]\,.
\end{aligned}
\label{eq:s3:semileptonicLeptonDecayRate}
\end{align}
The expression for decays to vector mesons vanishes at tree level because of the scalar nature of the interaction mediated by $a$, while for unflavored mesons, the situation is once again complicated by the overlap with $G\widetilde{G}$ \cite{Cornella:2019uxs}. Since the experimental constraints on the branching ratio of these decays are all quite similar, the most constraining ones are those involving decays of taus into muons \cite{ParticleDataGroup:2022pth} and all lead to comparable bounds $\vphi \gtrsim$ TeV.

For $\ell_i \rightarrow  \ell_j \ell_k \ell_k $, the branching ratio in the presence of an ALP is enhanced in the on-shell production regime $ 2m_k<m_a<m_i-m_j$. Here, the ratio can be factorized as
\begin{align}
    \text{Br}(\ell_i \rightarrow  \ell_i \bar \ell_k \ell_k) \simeq \text{Br}(\ell_i \rightarrow \ell_j a) \text{Br}(a\rightarrow \bar \ell_k \ell_k).
    \label{eq:s3:lepton3decay}
\end{align}
with relevant decay rates given in Eqs.~\eqref{eq:s3:Gammaaff} and \eqref{eq:s3:Gammafaf}. The situation for these decays is similar to that of semileptonic meson decays in that we can distinguish prompt and displaced vertex searches. Among the former, we find that the strongest constraint is due to $ \text{Br}(\tau \rightarrow3\mu) < 2.5 \times 10^{-8}$ at $95 \%$ CL \cite{Belle-II:2024sce} constraining up to $\vphi \gtrsim 10^5$ GeV. Decays into electrons are suppressed due to the hierarchical nature of $c^e$; even the strong bounds from $\mu\to 3e$ \cite{SINDRUM:1987nra} do not beat $\tau \rightarrow 3\mu$.
Searches for displaced vertices are instead lacking in this context, although new searches have been proposed over the last years~\cite{Heeck:2017xmg, Cheung:2021mol}. To our knowledge, the only available bounds are on $\mu\rightarrow 3e$ \cite{SINDRUM:1986klz} and $\mu \rightarrow e\gamma\gamma$ \cite{Bolton:1988af,Natori:2012gga, MEG:2020zxk}. These, however, lead to completely negligible bounds in our model of the order $\vphi \gtrsim $\,GeV.

\noindent
\emph{VLF}. The VLLs can induce the $\ell_i \to M^{(*)} \ell_j$ decays in multiple ways: either by a double insertion of the tree-level operators in Eqs.~\eqref{eq:s3:OHdOHu}, \eqref{eq:s3:OHl}, a single insertion of the 1-loop generated operator in Eq.~\eqref{eq:s3:Olldd}, or, for unflavored $M$, by a single LFV modification of the $Z$ boson couplings from Eq.~\eqref{eq:s3:OHl}. We find that the latter effect is phenomenologically the most important. Using \texttt{flavio}~\cite{Straub:2018kue} we obtain constraints from $\tau \to \rho \ell$ and $\tau \to \phi \ell$~\cite{Aebischer:2018iyb}, which are however of order $\vphi\gtrsim20~\mathrm{GeV}$ and hence subleading.

Also the $\ell \to 3\ell$ decays can proceed through a $Z$ exchange\footnote{It is worth mentioning that in our model the operator $\mathcal{O}_{\ell\ell} ^{ijkl} = (\bar \ell_i \gamma ^\mu \ell _j)(\bar \ell_k \gamma ^\mu \ell _l)$ is also generated at 1-loop integrating out the VLLs. This effect is, however, subleading compared to the tree-level one.} with the tree-level LFV coupling from Eq.~\eqref{eq:s3:OHl} and an SM $Z$ coupling. As the couplings in the VLL benchmark are anarchic, $\mu\to 3e$ is now the only phenomenologically important channel. The branching ratio prediction reads~\cite{Crivellin:2020ebi}
\begin{equation}
    \mathrm{Br}(\mu \to 3e) = \frac{m_\mu^5}{1536 \pi^3 m_Z^4 \Gamma_\mu} \left(2 |\Gamma_{e\mu}^{\ell L} \Gamma_{ee}^{\ell L}|^2 + |\Gamma_{e\mu}^{\ell L} \Gamma_{ee}^{\ell R}|^2  \right)\,,
\end{equation}
with $\Gamma_{e\mu}^{\ell L}$ defined in Eq.~\eqref{eq:S3:VLLcLFV} and with the SM $Z$ couplings to electrons. We obtain the bound $\vphi\gtrsim 400 \text{ GeV}$.

\subsubsection*{EDMs}

Electric dipole moments are among the most sensitive probes of CP violating new physics. The strongest constraints are the ones on the electron EDM, $|d_e|/e < 5.4 \times 10^{-30}$\,cm at $95\%$\,CL \cite{Roussy:2022cmp}, and on the neutron EDM, $|d_n|/e < 2.6 \times 10^{-26}$\,cm at $95\%$\,CL~\cite{Abel:2020pzs}.
In particular, the electron EDM at leading order is directly associated with the dipole operator $\mathcal{O}_{e\gamma}$ introduced earlier \cite{Kley:2021yhn}
\begin{align}
    d_e \simeq - 2 \Im  C_{e\gamma}^{11}\,,
\end{align}
while the expression for the neutron EDM, in the notation of \cite{Kley:2021yhn}, reads\footnote{We omit the contribution from $C_{Hud}$, since in any case below the EW scale this operator can be mapped to 4-fermion operators obtained integrating out $W^\pm$ bosons.}
\begin{align}
\begin{aligned}
    d_n \simeq& - 0.2 \, d_u + 0.8 \, d_d - 0.003 \, d_s + 0.05 \, \hat d_u+0.1 \, \hat d_d - 51 e \cdot \text{MeV} \, C_{G\widetilde{G}} 
    \\
&- 0.6  e \times \text{GeV} 
\Im \left[ C_{ud} ^{S1,RR} + C_{ud} ^{S8,RR} - C_{duud} ^{S1,RR} -C_{duud} ^{S8,RR}\right ]^{1111}.
\label{eq:s3:nEDM}
\end{aligned}
\end{align}
In principle, our model also predicts contributions to the magnetic dipole moments, which are much less constraining than those from the EDMs.

\noindent
\emph{ALP}. In the heavy ALP case ($m_a \gtrsim$ GeV), the leading contribution to the eEDM can be directly read off from Eq.~\eqref{eq:s3:CegammaALP}. The imaginary part of $C_{e\gamma} ^{11}$ vanishes, meaning that no contribution to the eEDM is generated at this order. To check that NLO corrections do not lead to signification bounds, we can estimate the hypothetical constraint if an $\Oone$ imaginary part in $C_{e\gamma}$ was there, obtaining the weak bound $\vphi > \mathcal{O}(10) \text{ GeV} \times (\text{GeV}/m_a)$, indicating that this observable is under control. For the neutron EDM, the situation is similar: the dipoles ($d_u, d_d, d_s, \hat d_u, \hat d_d$) feature an expression close to that of $C_{e\gamma}$ and are hence real. The 4-fermion operator coefficient $C_{ud}^{S1,RR}$ entering Eq.~\eqref{eq:s3:nEDM} is generated at the tree level, but features no imaginary part. An analysis analogous to the one for the eEDM reveals that the bound for a hypothetical $\Oone$ phase reads $\vphi > \mathcal{O}(10^2) \text{ GeV} \times (\text{GeV}/m_a)$, once again negligible compared to the flavor bounds analyzed earlier. In the very light ALP case ($m_a \lesssim $ GeV), contributions to the nEDM approximate to a constant close to the previous estimate when $m_a \sim m_u$ \cite{DiLuzio:2020oah}. While this could potentially lead to a bound comparable to the ones from other observables analyzed earlier, in reality, we know that this contribution is still not there due to the absence of an imaginary part in the relevant Wilson coefficients. Therefore, whatever effect may arise at NLO must be phenomenologically irrelevant.
The operator $C_{G\widetilde{G}}$ can be generated only at very high loop orders. 

\noindent
\emph{VLF}. The VLFs generate the operators Eq.~\eqref{eq:s3:OeBPhi} directly contributing to the eEDM. We can estimate the size of the bound employing Eq.~\eqref{eq:s3:CeBWPhiMatchete}, that after $e_i$ rotation generates a non-zero entry $C_{e\gamma}^{11} \sim e  v y_*^5 \eps^4/16\pi^2 \vphi^2$ with an $\Oone$ imaginary part. This is a single contribution to the eEDM, which, in principle, should be added to that of the dimension-8 $C_{eB\Phi, eW\Phi}^{i,1}$, whose induced $C_{e\gamma}^{11}$ features the same scaling. 
However, barring accidental cancellations, this should be enough to estimate the constraint coming from the eEDM. We find the bound $\vphi \gtrsim 230$ GeV for our benchmark. This is interesting but again subleading compared to $\eps_K$ and other bounds derived earlier.
For the nEDM, the bound from the dipole operators can be obtained in an analogous way to that of the eEDM. As the experimental bound on the former is 4 orders of magnitude weaker than the one on the latter, and given the same functional dependence on the UV scales and couplings (but in the quark sector), this does not lead to a competitive bound on $\vphi$. The 4-fermion operators entering Eq.~\eqref{eq:s3:nEDM} are associated to $\mathcal{O}_{quqd}$ in the SMEFT, but in the SM+$\Phi$ EFT are uplifted to $\mathcal{O}_{quqd\Phi} ^{ijkl} \sim \Phi^{n_{ik}^q} (\bar q_i u_j) \eps (\bar q_k d_l)$. Due to the large $\eps$ suppression, this does not lead to relevant constraints. Finally, the full expression for Eq.~\eqref{eq:s3:nEDM} should include also the contribution from $C_{Hud} ^{ij} = - (y_i ^{u a})^* y_j ^{da}/M_{Q_2}^2$ \cite{Kley:2021yhn}. We find $\Im C_{Hud} ^{11} =0$; hence this cannot contribute to leading order. A hypothetical $\Oone$ imaginary part would, in any case, result in the subleading $\vphi > |y_*| \times \mathcal{O}(10^2)$ GeV, hence NLO effects on this operator can be safely neglected. The Weinberg operator is generated only at a very high loop order.

In conclusion, the current constraints on EDMs generate only subleading bounds compared to the other observables we analyzed.

\subsubsection*{Cosmology, astrophysics, and haloscopes}

\begin{figure}[t]
    \centering
    \includegraphics[width=0.9\textwidth]{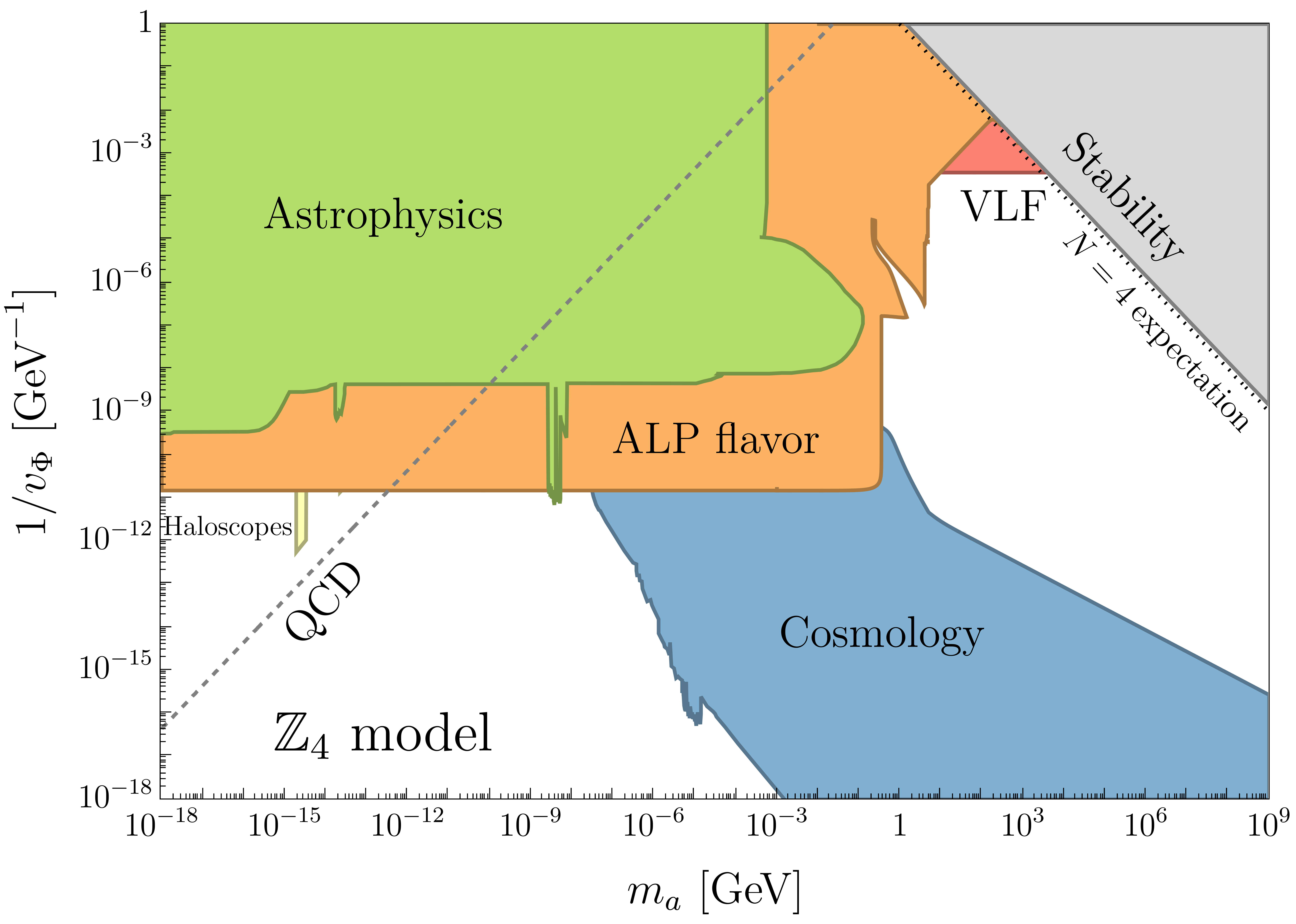}
    \caption{A global overview of the phenomenology of a FN ALP in the $(m_a, 1/\vphi)$ plane. The gray region is excluded by the stability of the potential, the dotted line shows the $N=4$ expectation, and the dashed line shows the QCD expectation, as discussed around Figure~\ref{fig:s2:theoryPlot}. The rest of the contours are due to the constraints discussed in detail in Section~\ref{sec:z4observables}. The red contour is due to the constraints from the UV VLFs, the orange contour is due to the combined constraints on the FN ALP from flavor observables, while the green, yellow, and blue contours are due to astrophysics, haloscopes, and cosmology, respectively. See the surrounding text for more details.}
    \label{fig:Z4plot_big}
\end{figure}

We mention the impact of haloscope searches, astrophysics, and cosmology bounds on our model for completeness. Haloscope searches constrain $\vphi$ through the axion-photon coupling. We take the bounds from \cite{AxionLimits}, rescaling the plot to adapt it to our model correctly. The usual astrophysics bounds (from \cite{AxionLimits}) are completely superseded by the bound due to $K\rightarrow \pi a$ and hence are irrelevant. For cosmology, the left part of the blue region in Fig.~\ref{fig:Z4plot_big} is taken again from \cite{AxionLimits} and is dominated by the $X$-rays bound, assuming that the ALP constitutes a significant portion of the dark matter observed in our universe. The rightmost part is obtained via the requirement that the total ALP decay rate satisfies $\Gamma_\text{a,tot} \geq 3 H(T_\text{BBN})$, where $T_\text{BBN} = 4$ MeV, in order to comply with the constraints from Big Bang Nucleosynthesis \cite{deSalas:2015glj}. The rate $\Gamma_\text{a,tot}$ has been computed taking into account all the decay channels of the ALP in our model. For ALP masses around the QCD confinement scale, this has been approximated with a linear interpolation between $m_a \sim 3m_\pi$ and $m_a \simeq$ 2 GeV, given the complications due to flavor and CP violating ALP interactions with hadrons present in our model. The impact of the heavy $\rho$ and the VLFs is irrelevant as they quickly decay to SM particles via Eqs.~\eqref{eq:s3:L_VLFs} and \eqref{eq:s3:rhoa_couplings}.

\subsubsection*{Direct searches}

Low-scale FN models present an exciting opportunity for direct searches at high-energy colliders. Collider searches for heavy neutral scalars and ALPs have recently received significant attention~\cite{Mimasu:2014nea, Brivio:2017ije,Carmona:2022jid, Esser:2023fdo, Blasi:2023hvb,Rygaard:2023dlx, Phan:2023dqw, Biswas:2023ksj, Li:2024thq}.
Light ALPs ($m_a\lesssim 10$ GeV) either escape the detector or lead to displaced vertices.
Mono$-\gamma$ and mono-jet searches for light ALPs \cite{Mimasu:2014nea} are, however, not competitive with meson decays. Moving on to EW scale masses, phenomonology in this model is driven by couplings to second-generation fermions, see Eqs.~\eqref{eq:s3:cdmatrix} and \eqref{eq:s3:cecuMat}. This is because, in the $\Z_4$ case, the bottom and tau Yukawas size are not dynamically explained (unlike in the $\Z_8$ model of Section~\ref{sec:Z8}). Resonant axion production at hadron colliders has been discussed in Ref.~\cite{Bauer:2016rxs}. We expect the dominant channels in our case to be $gc\rightarrow ta$ and $\bar c c\rightarrow a$. The cross sections, computed in Fig.~11 of Ref.~\cite{Bauer:2016rxs} for $\vphi = 0.5$ TeV, show that one should expect a large number of events. However, as discussed in \cite{Bauer:2016rxs}, due to large backgrounds a dedicated detailed experimental analysis is required.

The flavor-violating ALP-top-charm coupling leads to the $t\rightarrow c a$ decay. A study of this anomalous decay has been carried out in~\cite{Bauer:2016rxs}, considering the current and the potential future bounds on $\text{Br}(t\rightarrow H c)$ with $H\rightarrow \bar b b$ and assuming $\text{Br}(a\rightarrow \bar b b)>80\%$. A major difference between ours and their scenario is the branching ratio of the decay $a \rightarrow \bar b b$. This would reduce the impact of this search since we expect $a \to c \bar{c}$ decay channel to dominate. A proper assessment would require repeating the numerical study from~\cite{Bauer:2016rxs} within our model, which is beyond the scope of our paper. Therefore, we will not include these bounds in our figures. Even considering the nominal value they find, the bound from $\Delta F=2$ processes remains dominant in that region. However, with future measurements, the situation might change. For light ALPs, the displaced signatures from top decays could lead to important bounds in the future \cite{Carmona:2022jid}.

For completeness, searches for vectorlike quarks and leptons at LHC set lower bounds on their masses of $\mathcal{O}$(TeV)~\cite{ATLAS:2024gyc, CMS:2024bni}, which cannot compete against bounds from flavor physics in our scenario, forcing $M \approx \vphi/ \eps \gtrsim $ PeV.

To sum up, while the current direct searches provide subleading constraints, the situation will improve with proposed future high-energy colliders.

\subsection{Summary of the results}
\label{sec:z4results}

This subsection summarizes and illustrates the main phenomenological results derived in the previous subsection. In Fig.~\ref{fig:Z4plot_big}, we show the exclusion contours in the $(m_a, 1/\vphi)$ plane for a broad range of ALP masses, $m_a \in [10^{-18}, 10^9]~\mathrm{GeV}$. The stability region in gray, the dotted line for the $N=4$ expectation, and the dashed line for the QCD expectation are the same as already discussed in Fig.~\ref{fig:s2:theoryPlot}. Importantly, we can immediately notice that following the $N=4$ expectation line, the main constraint on the $\Z_4$ model is due to the vector-like fermions, illustrating the significance of a UV completion. The contour in red depicts the combined constraint due to the presence of these states in the UV, which is dominated by the CP violating kaon mixing observable $\epsilon_K$ as discussed in the previous subsection, setting $\vphi \gtrsim 2.8~\mathrm{TeV}$. The answer to the question posed in the introduction, ``How low can the FN symmetry-breaking scale be?", aligns with the current energy frontier. 

The ALP flavor phenomenology is less important in this particular scenario, even though it plays a major role in a large part of the parameter space as depicted by the orange contour, which combines all of the flavorful ALP constraints from the previous subsection. Notably, it sets the leading constraint on the flaxion/axiflavon scenario~\cite{Calibbi:2016hwq,Ema:2016ops} through the kinematically allowed $K\to \pi a$ decays. As already mentioned, the astrophysics constraints depicted in green are less important in scenarios with flavorful ALPs. In blue, we show the constraints from cosmology mentioned earlier, with the left part taken from Ref.~\cite{AxionLimits} and the right coming from the requirement of not spoiling the success of BBN. Although these constraints are important in a large part of the parameter space, we now \textit{zoom in} to a region that is arguably more interesting for low-scale FN models.

\begin{figure}[t]
    \centering
    \includegraphics[width=0.9\textwidth]{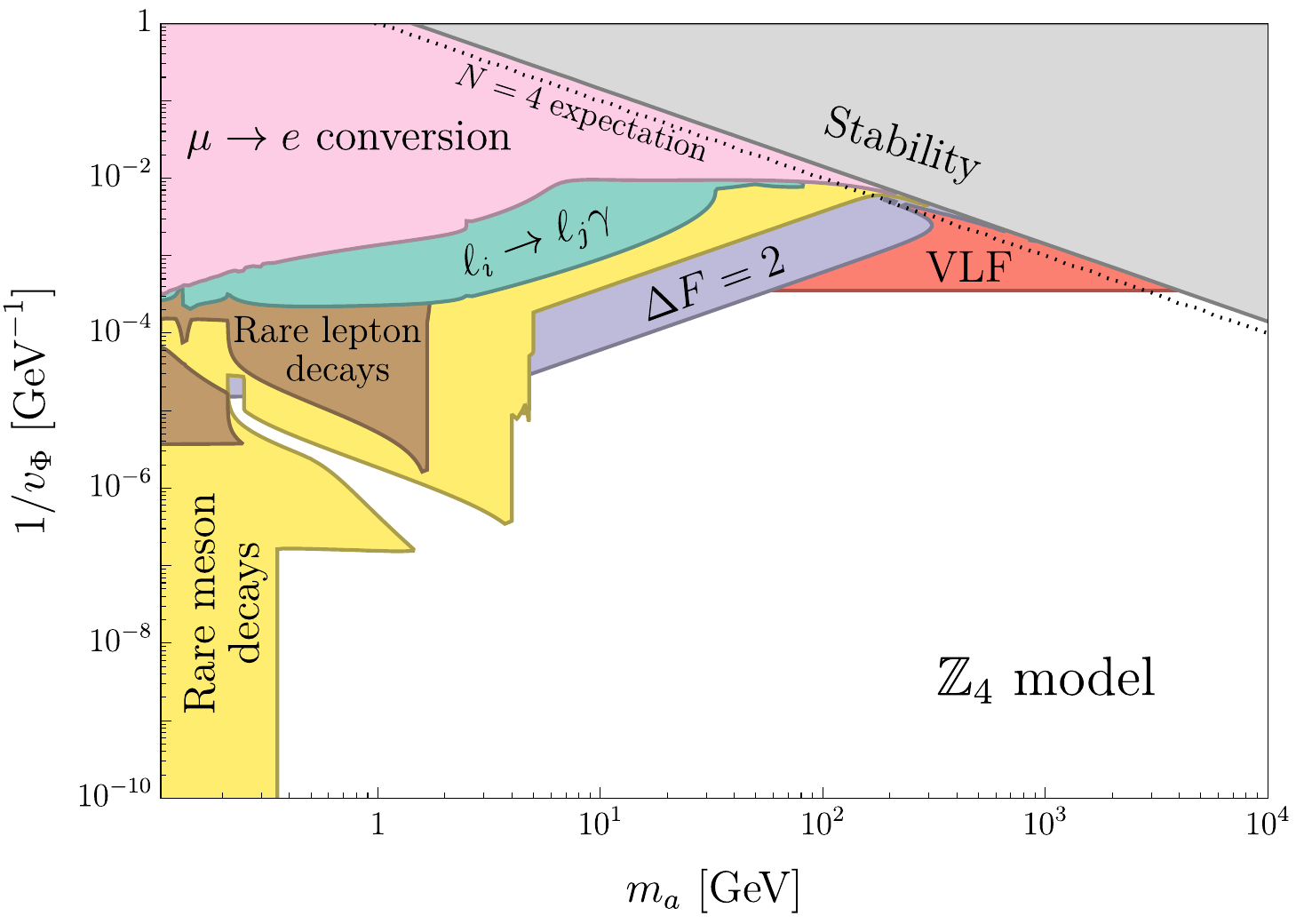}
    \caption{A zoomed-in version of Figure~\ref{fig:Z4plot_big}, focusing on the parameter space interesting for low-scale FN models. Again, the gray region is excluded by potential stability, while the dotted line shows the expectation of $N=4$. We show the leading constraint from VLFs in red and a compendium of constraints on the flavorful ALP in the rest of the contours. See the surrounding text for a detailed discussion.}
    \label{fig:Z4plot_zoomed}
\end{figure}

In Fig.~\ref{fig:Z4plot_zoomed}, we show a zoomed-in version of Fig.~\ref{fig:Z4plot_big}, now focusing on the ALP masses in the range $m_a \in [0.1, 10^4]~\mathrm{GeV}$. This region of the parameter space is of main interest for the FN models studied in this work and is constrained mainly from flavor physics observables. We emphasize again that the constraint due to VLFs dominates the natural $\Z_4$ model. However, by allowing for $\lambda_N^\prime \ll 1$ we can in principle populate the whole parameter space. Moreover, as we will see in the next section, the qualitative picture varies little, even for more elaborate $\Z_N$ models. Since the ALP flavor constraints dominate a large portion of this parameter space, we now split them into multiple contours to differentiate between observables. In the subsequent discussion, we move from high to low ALP masses, commenting on the leading constraints for the different regimes. 

In purple, we show the combined constraints from the ALP contributions to neutral meson mixing observables, here dominated by $\epsilon_K$. These provide the leading constraint for $m_B \lesssim m_a \lesssim 60~\mathrm{GeV}$, setting $\vphi \gtrsim 10 - 100~\mathrm{TeV}$. Rare meson decays become kinematically allowed for ALP masses below $m_B$, as depicted in the yellow contour. At first, $B$-meson decays to visible final states provide the leading constraint, dominated by displaced vertex analyses of $B\to K^{(\ast)} \mu \mu$ by LHCb, setting $\vphi \gtrsim 100-1000~\mathrm{TeV}$. The region below that is due to $B\to K a$, setting an upper and a lower bound on $\vphi$ as we demand the ALP to behave as an invisible final state, as discussed in the previous subsection. Finally, for ALP masses below $m_K$, $K\to \pi a$ decays completely dominate the phenomenology. 

For completeness we comment on the subleading LFV phenomenology: in brown contours, we depict the combined constraints from rare lepton decays, dominated first by $\tau \to 3\mu$ and then by $\tau \to \mu +\mathrm{invisible}$, with the flat region being associated to the plateau of $\tau \to M^{(\ast)} \mu $ where $M^{(\ast)}$ are unflavored mesons. In water green, we show the constraints from $\ell_i \to \ell_j \gamma$, with the sensitivity of up to $\vphi \gtrsim 10~\mathrm{TeV}$. Finally, in pink, we show the contour from $\mu \to e$ conversion, currently setting bounds up to $\vphi \gtrsim $ TeV.

Looking ahead, several experimental advancements will significantly probe the allowed parameter space, see e.g.~\cite{Calibbi:2020jvd, Jho:2022snj}. First, future bounds on charged lepton flavor violation from $\mu$ to $e$ conversion~\cite{Bernstein:2019fyh, Moritsu:2022lem} will impose stricter constraints on the UV sector, pushing the FN scale by another order of magnitude. Second, there is a pressing need for dedicated experimental programs to search for many $M \to M' a$ and $\ell \to \ell' a$ processes with displaced ALP vertices $a \to \ell_i \ell_j, \gamma \gamma$, etc. Finally, both Belle-II and LHCb will improve the existing limits on rare decays, further pushing into the unexplored territory.

\section{A finer model: $\Z_8$}
\label{sec:Z8}

The simplicity of the $\Z_4$ model comes at the expense of not perfectly explaining the quark and lepton flavor hierarchies. While the existence of hierarchies between generations is consistently predicted, this model fails to account for the smallness of the bottom and tau Yukawas, which are observed to be $\mathcal{O}(0.01)$ instead of the expected $\mathcal{O}(1)$. Furthermore, the mass hierarchy between different generations of down quarks and charged leptons is more compressed compared to that of the up quarks, despite the $\Z_4$ model predicting a universal $\epsilon$ factor between generations for all three gauge representations. These additional finer features of the SM flavor puzzle call for a larger symmetry.

\paragraph{$\Z_8$ EFT.} The up-quark sector exhibits a double hierarchy compared to the down sector. This feature was recently successfully achieved by the $\U(2)_{q+e^c+u^c} \times \Z_2$ symmetry in Ref.~\cite{Antusch:2023shi}, which retains the favorable characteristics of $\U(2)_{q+e^c}$, including the neutrino anarchy, while producing a single (double) hierarchy in the down (up) quarks. The $\Z_2$ factor serves to suppress overall the down sector, which explains the smallness of $y_{b,\tau}$. This section demonstrates that the smallest discreet symmetry capable of reproducing the same texture as in Ref.~\cite{Antusch:2023shi} is $\Z_8$.

To illustrate this, if the hierarchy parameter in the down sector is $\epsilon$, then in the up sector, it should be $\epsilon^2$, leading to $y_u / y_t \sim (y_d / y_b)^2 \sim \epsilon^4$. Following the discussion below Eq.~\eqref{eq:s2:YukLagr}, given the largest $n^f_{ij} = 4$, the minimal symmetry group required is therefore $\Z_8$. The non-trivially charged fields in this case are $[q_1] = -[e_3] = -[u_1] = 2$, $[q_2] = -[e_2] = -[u_2] = 1$, $[d_i] = -2$ and $[\ell_i] = 4$, where $i=1,2,3$. The leading patterns are
\begin{align}
\label{eq:s4:Z8nij}
    n_{ij} ^{d} =
    \begin{pmatrix}
        4 & 4 & 4\\
        3 & 3 & 3\\
        2 & 2 & 2
    \end{pmatrix},
    \quad
    n_{ij} ^{e \prime}=
    \begin{pmatrix}
        4 & 3 & 2\\
        4 & 3 & 2\\
        4 & 3 & 2
    \end{pmatrix},
    \quad
    n_{ij} ^{u} =
    \begin{pmatrix}
        4 & 3 & 2\\
        3 & 2 & 1\\
        2 & 1 & 0
    \end{pmatrix}
,
\end{align}
while $n_{ij} ^{d \prime}$, $n_{ij} ^{e}$ and $n_{ij} ^{u \prime}$ give comparable or subleading contributions to $Y^f_{ij}$. 

A few comments are in order. 
First, we can see explicitly why the $\Z_8$ symmetry is necessary. A replacement $\Phi^M \rightarrow (\Phi^*)^{N-M}$ would disrupt the hierarchies if $N < 8$. In addition, the universal charges for $d_i$ and $\ell_i$ are chosen to be different, even though they lead to the same singular values $\hat{y}^{d,e}_{ii} \sim \{ \epsilon^4, \epsilon^3, \epsilon^2 \}$. This choice allows the CKM matrix to be approximately a unit matrix while implying no selection rules on the neutrino sector. In other words, the Weinberg operator
\begin{equation}
-\mathcal{L} \supset \frac{y^\nu_{ij}}{M_\nu} \ell_i \ell_j H H \,,
\end{equation}
is neutral under the $\Z_8$ symmetry. Therefore, it features anarchic coefficients, which is consistent with the large observed mixings in the PMNS matrix.\footnote{In the type-I seesaw scenario, completing this operator with heavy right-handed neutrinos $\nu_i$ means $[\nu_i] = 4$, ensuring that both the Yukawa interaction and Majorana mass terms are allowed by the $\Z_8$ symmetry.}

The details of fitting $x^f_{ij}$ and $x^{f\prime}_{ij}$ from Eq.~\eqref{eq:s2:YukLagr} to the observed SM flavor parameters are presented in Appendix~\ref{sec:app:z8}. As demonstrated by the numerical benchmark point, this FN EFT offers an excellent description of the SM flavor parameters for $\epsilon \simeq 6.6 \times 10^{-2}$ with all fit parameters being $\mathcal{O}(1)$, as anticipated.

\paragraph{UV completion.} The set of VLF with masses at the scale $M$, completing Eq.~\eqref{eq:s2:YukLagr} for the $\Z_8$ charges listed above, is somewhat more complex than in the previous section. Let us start with the quark sector. The charges of the new VLF fields are
\begin{equation}
\begin{split}
[Q^a_2]&=0,\quad [Q_1]=1,\\
[U^a_2]&=0,\quad [U_1]=-1,\\
[D^i_2]&=0,\quad [D^i_1]=-1.
\end{split}
\end{equation}
The gauge representations correspond to the SM $q$, $u$, and $d$ fields, respectively, with $a=1,2$ and $i=1,2,3$. Adding these states in the UV allows for the tree-level completion of the effective quark Yukawa operators. The corresponding diagrams are analogous to those in Fig.~\ref{fig:diag}, featuring $n^f_{ij}$ VLF propagators. The effective Yukawa matrix entries parametrically scale as
\begin{equation}
    Y^f_{i j} \sim (y_{*})^{n^f_{i j}+1} \epsilon^{n^f_{i j}}.
\end{equation}
Here, $y_{*}$ generically represents a dimension-4 coupling in the UV Lagrangian. As previously discussed, the product of several $\mathcal{O}(0.3)$ couplings can contribute to hierarchies. Consequently, we expect the variance in the UV couplings to be even smaller than in the EFT fit presented above, similar to what was observed for the $\Z_4$ model.

Let us now turn our discussion to the charged lepton sector. The required set of VLF, all in the gauge representation of the SM field $\ell$, is
\begin{equation}\label{eq:sec4:leps}
    [L_N]=0,\quad [L^a_{N-1}]= -1, \quad [L^i_{N-2}]= -2, \quad [L^i_{N-3}]= -3,
\end{equation}
where $N=8$. Integrating them out at the tree level produces the desired charged lepton effective Yukawa operators. An example diagram involves connecting $e_1$ (which has charge zero) with $\ell_i$ (charge 4) through the aforementioned chain of VLF, attaching $H$, and subsequently adding four $\Phi^*$.

The final model-building task is to endow the ALP with a non-null mass. The simplest option is to consider the wheel diagram already introduced in Fig.~\ref{fig:s2:wheel}, which requires a full set of VLF representation charges. Since we already have half of the $L_k$ charges, we only need, in addition,\footnote{Due to the large number of $\SU(2)_{\rm L}$ doublets in this model, the gauge coupling $g_2$ develops a Landau pole at approximately $10^{14}$\,GeV for $M=1$\,PeV (1-loop estimate). In contrast, for the $\Z_4$ model, the couplings remain perturbative up to the Planck scale.} 
\begin{equation}
    [L_1]=1,\quad [L_{2}]= 2, \quad [L_{3}]= 3, \quad [L_{4}]= 4,
\end{equation}
which together with Eq.~\eqref{eq:sec4:leps} gives the diagram in Fig.~\ref{fig:s2:wheel}.
The prediction for the ALP mass aligns precisely with the one discussed in Section~\ref{sec:theory} for the $N=8$ case, as shown in Fig.~\ref{fig:s2:theoryPlot}. We expect $m_a / v_\Phi \sim \{10^{-6}, 10^{-4}\}$, see Eq.~\eqref{eq:s2:wheelma} and the discussion below it. Thus, unlike in the $\Z_4$ model, here we expect a light ALP, potentially leading to interesting effects described by the SM$+a$ EFT.

\paragraph{Phenomenology.} 
First, let us discuss the UV contributions to operators without $\Phi$ insertions before moving on to the ALP.

\emph{VLF}. The leading effect arises from $\mathcal{L} \supset y^{ij}_{D_1} \bar{D}^i_1 \Phi d^j$, which, after integrating out $D^i_1$ at the one-loop level, matches to the $(\bar{d}_1 \gamma_\mu d_2)^2$ operator. This is analogous to the $\Z_4$ model, where the $H$ and $Q^1_2$ fields appeared in the box diagram, whereas in this case, we have the $\Phi$ and $D_1^i$ fields (see Eq.~\eqref{eq:s3:Odddd&Ouuuu} and the subsequent discussion). Therefore, the leading constraint comes from $\epsilon_K$ and can be easily derived by rescaling the $\Z_4$ case by the $\epsilon$ ratios in the two models. Since $\epsilon$ is larger in the $\Z_8$ model, the bound becomes $v_\Phi \gtrsim 40$\,TeV, up to $\mathcal{O}(1)$ variation in $y^{ij}_{D_1}$. 

\begin{figure}[t]
    \centering
    \includegraphics[width=0.9\textwidth]{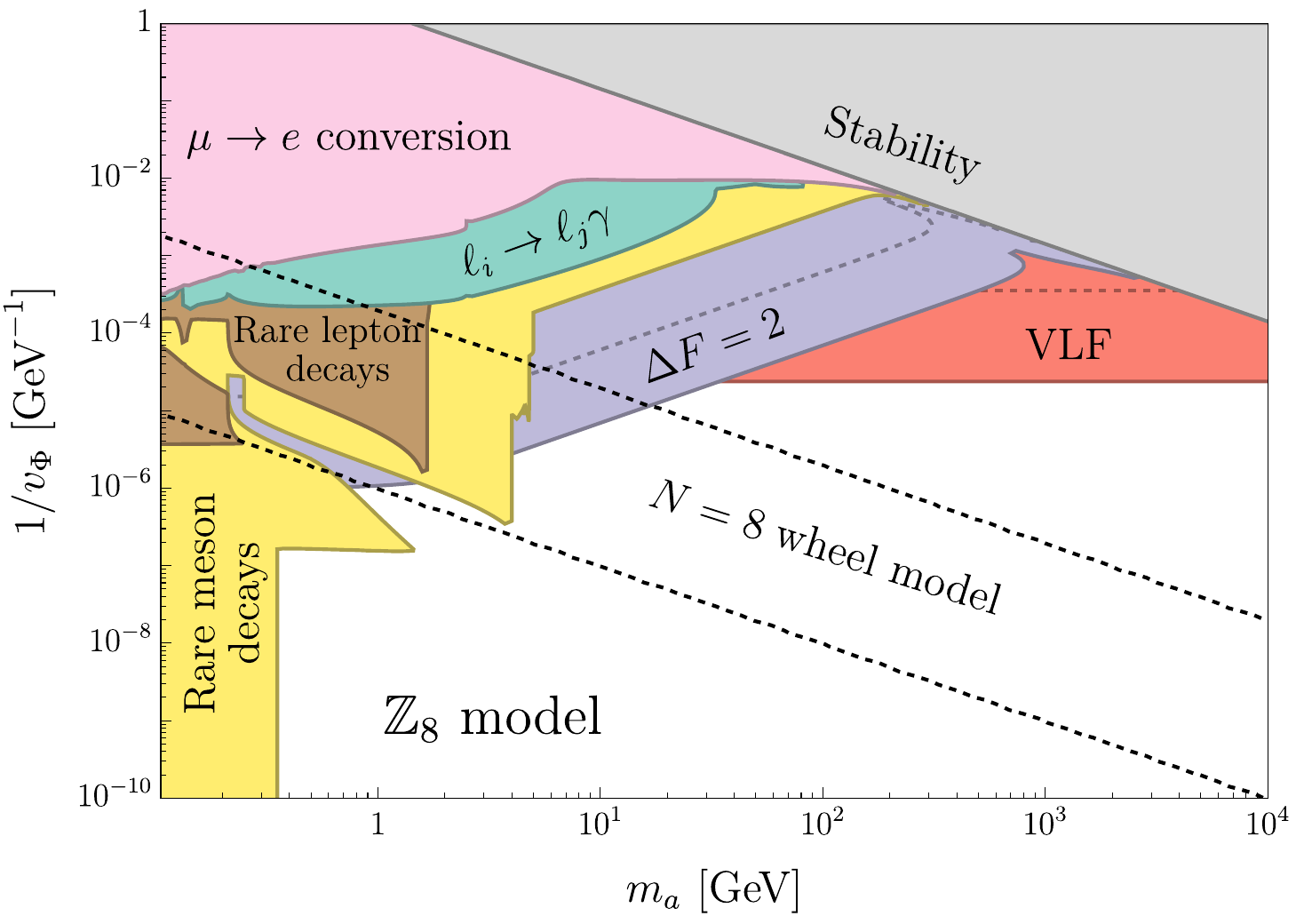}
    \caption{Constraints in the $(m_a, 1/\vphi)$ plane, similar to Figure~\ref{fig:Z4plot_zoomed}, but now for the $\Z_8$ model. The gray region is excluded by vacuum stability, while the dashed black lines show the expected range in a $N=8$ wheel model. The red and purple (full) contours show the major differences with respect to the $\Z_4$ case (dashed). The rest of the constraints are qualitatively the same as in the $\Z_4$ case. See the surrounding text for more details.}
    \label{fig:Z8plot_zoomed}
\end{figure}

\emph{ALP}. Assuming $v_\Phi$ saturates this bound, we expect $m_a \in (0.04, 4)$\,GeV. This mass range coincides with numerous significant flavor constraints in the $\Z_4$ model, providing leading bounds on the FN scale. We first need to examine the ALP coupling matrices to compare the predictions between the two models. In this case, the hierarchies in the left-handed rotation matrices $U_L^d$ and $U_L^u$ are similar. Hence, the CKM receives, in principle, a sizeable contribution from both.
Furthermore, this case has many more free parameters for $Y_u$. For this reason, we find it more instructive to report the general structure of the couplings in the $\Z_8$ model up to $\mathcal{O}(1)$ factors:
\begin{align}
\begin{gathered}
    c^d  \sim 
    \begin{pmatrix}
        m_d  & m_s \frac{m_d}{m_s}  & m_b \frac{m_d}{m_b}\\
        m_d  \frac{m_d}{m_s}  & m_s &  m_b \frac{m_s}{m_b} \\
         m_d \frac{m_d}{m_b} &  m_s \frac{m_s}{m_b} & m_b\\
    \end{pmatrix}, \qquad
    c^u \sim
    \begin{pmatrix}
        m_u  & m_c \sqrt{\frac{m_u}{m_c}} & m_t \sqrt{\frac{m_u}{m_t}} \\
        m_c \sqrt{\frac{m_u}{m_c}} & m_c & m_t \sqrt{\frac{m_c}{m_t}} \\
        m_t \sqrt{\frac{m_u}{m_t}} & m_t \sqrt{\frac{m_c}{m_t}} & m_t \frac{m_c}{m_t}\\
    \end{pmatrix},
    \\
    c^e \sim
    \begin{pmatrix}
        m_e  & m_e \frac{m_e}{m_\mu}  & m_e \frac{m_e}{m_\tau}\\
        m_\mu  \frac{m_e}{m_\mu}  & m_\mu &  m_\mu \frac{m_\mu}{m_\tau} \\
         m_\tau \frac{m_e}{m_\tau} &  m_\tau \frac{m_\mu}{m_\tau} & m_\tau\\
    \end{pmatrix}.
\end{gathered}
\end{align}
The most prominent differences compared to the $\Z_4$ case are $c^{e}_{33}, c^d _{33}$, which now are unsuppressed due to the fact that the third generation is charged, and of course, the whole $c^u$. The fact that $c^u$ mixings scale as the square root of the ratios of masses is due to the fact that neither $q_i$ nor $u_i$ are ``universally'' charged, unlike $d_i$ and $e_i$. In comparison with the latter two, the up sector has doubled suppression. Hence, the mixings are ``split'' between the left and right rotation matrices, which give comparable contributions to $c^u$, see Eq.~\eqref{eq:appB:cformula}. Because of this, ALP contributions to $D-\bar{D}$ mixing become now more important \cite{Carmona:2021seb}.
In particular, if the CKM matrix is still fully given by $U_L^d$, as is for simplicity assumed in the example of App.~\ref{sec:app:z8}, the constraint from $\Z_4$ due to $\epsilon_K$ remains essentially the same. Hence, in this case, CP violation in D meson mixing is actually the most important $\Delta F=2$ probe. In the general case in which the CKM receives sizeable contributions from both the diagonalizing matrices, we find that the $\epsilon_K$ bound becomes slightly stronger than in the $\Z_4$ case, leading to bounds comparable to that from CP violation in $D-\bar{D}$ mixing.

We illustrate the phenomenology of the $\Z_8$ model in Figure~\ref{fig:Z8plot_zoomed}. The dashed black lines now show the expectation in the $(m_a, 1/\vphi$) plane for a $N=8$ wheel model, as already discussed around Figure~\ref{fig:s2:theoryPlot}. With dashed red and purple lines we show the $\Delta F=2$ and VLF constraints in the case of $\Z_4$, as in Figure~\ref{fig:Z4plot_zoomed}. The full contours show the respective bounds in the $\Z_8$ case, both stronger in line with the discussion above. It is interesting to note how in this scenario, contrary to the case of $\Z_4$, the leading phenomenology is due to the light flavorful ALP and no longer due to the VLFs. In particular, the $\Delta F=2$ constraints together with displaced vertex searches in $B\to K^{(*)} \mu \mu$ (quantitatively the same as in $\Z_4$), set the most stringent constraints of order $\vphi \gtrsim 10^5 - 10^7 ~\mathrm{GeV}$ for the corresponding expected mass range of $m_a \in [1, 20]~\mathrm{GeV}$.

\section{Conclusion}
\label{sec:conc}

The presence of an ALP in quark and lepton flavor physics opens up a diverse array of new phenomena, making it an exciting area of research for phenomenologists. The appropriate theoretical framework to study these effects is the SM+$a$ EFT, which constructs a tower of local operators that respect both the SM symmetry and the approximate ALP shift symmetry. However, depending on the coefficients of the operators and the ALP mass, which are input parameters a priori, the expected patterns of deviations vary widely, thereby reducing the predictive power of this approach.

In the bulk of this paper, we surveyed the flavor physics phenomenology in the presence of an ALP, adding, however, a deeper UV context to the study. Specifically, we imposed a discrete FN symmetry motivated by the SM flavor puzzle. This symmetry provides power counting and selection rules that correlate various observables, offering a structured and focused framework for studying ALP effects in flavor physics. The main results are summarized in Figures~\ref{fig:Z4plot_zoomed} and~\ref{fig:Z8plot_zoomed}, which depict the parameter space of the ALP mass versus the FN scale. The two plots interpret a plethora of flavor physics measurements discussed in Section~\ref{sec:z4observables}, organized into suitable categories, within two concrete models based on $\Z_4$ and $\Z_8$ FN symmetries, respectively. The two choices allow us to quantify the uncertainty due to variations in the realization of the FN mechanism.

Prior to the comprehensive phenomenological work, we introduced a whole class of FN models based on $\Z_N$ symmetries. While a commonly considered perturbatively-exact $\U(1)_{\FN}$ predicts a very light (pseudo) Goldstone boson, the advantage of the discrete symmetries is that $m_a$ can be considerably heavier, opening up new phenomenological possibilities as discussed above. We began in Section~\ref{sec:theory} by charting the ALP parameter space based on theoretical considerations, as summarized in Fig.~\ref{fig:s2:theoryPlot}, and by presenting concrete predictive models, such as the wheel model shown in Fig.~\ref{fig:s2:wheel}.  This setup is more predictive than an explicitly broken $\U(1)_{\FN}$, which can produce correct textures provided $m_a \lesssim v_\Phi$ but does not correlate the two parameters. Nevertheless, our results qualitatively apply to the latter case as well. Assuming marginal couplings of $\mathcal{O}(1)$, our models predict a region in $(m_a, v_\Phi)$ as illustrated in the plots for the two chosen examples. Notably, the $\Z_4$ model is unique in predicting $m_a \sim v_\Phi$ due to the presence of a $\U(1)_{\FN}$-breaking dimension-4 term in the potential.

Another advantage of the $\Z_N$ class is the ability to balance between simplicity versus ``baroqueness", controlled by the value of $N$. Unsurprisingly, a large enough $N$ allows for a perfect fit of the SM flavor parameters. However, we prioritized simplicity, accepting an increased variance in the UV parameters. Following this path, we uncovered a remarkably simple and elegant model based on the $\Z_4$ group (Section~\ref{sec:z4setup}). By charging only left-handed quarks and right-handed charged leptons, this model generates hierarchies in the charged fermion masses and left-handed quark mixing matrices while maintaining an anarchic neutrino flavor structure. This minimal symmetry framework produces similar Yukawa textures as the full $\U(2)_{q+e}$ symmetry recently proposed in~\cite{Antusch:2023shi}. To address subtleties of the SM flavor puzzle, such as the smallness of $y_{b,\tau}$ and the difference in up versus down quark mass hierarchies, we introduced a refined $\Z_8$ model in Section~\ref{sec:Z8}. This model is carefully crafted, drawing inspiration from the novel $\U(2)_{q+e^c+u^c} \times \Z_2$ symmetry proposed recently in~\cite{Antusch:2023shi}.

A crucial aspect of our work is that in both scenarios, we constructed explicit renormalizable completions featuring new vector-like fermions with masses above the FN scale, given by $v_\Phi / \epsilon$ where $\epsilon$ is the FN spurion controlling the Yukawa textures. These new fields generate effective Yukawa operators in Eq.~\eqref{eq:s2:YukLagr}, but can also lead to deviations in rare flavor-changing neutral currents. An intriguing question arises about how the constraints on the minimal UV sector, which is essential for completing the FN mechanism, compare to those from a light ALP. The explicit models provide a basis for conducting precision calculations and studying this interplay between the ALP and the irreducible UV-induced effects. The so-called VLF bounds, shown in red in Figures~\ref{fig:Z4plot_zoomed} and \ref{fig:Z8plot_zoomed}, are dominant for a heavy ALP ($m_a \gtrsim 100$\,GeV). This parameter space region permits the lowest FN scale $v_\Phi$. For instance, in the case of $\Z_4$ predicting $m_a \sim v_\phi$, the FN symmetry could be restored even at the TeV scale. 

In conclusion, with its rich phenomenology and solid theoretical foundations, the FN ALP represents an exciting physics frontier for ongoing and future experiments in flavor and collider physics.

\acknowledgments

This work received funding from the Swiss National Science Foundation (SNF) through the Eccellenza Professorial Fellowship, project number 186866, ``Flavor Physics at the High Energy Frontier.''

\appendix

\section{UV and EFT parameters matching}
\label{app:paramsMatching}

\subsection{$\Z_4$ model}

\subsubsection{EFT}

In the $\Z_4$ model, other than $\Phi$, only $q_i$ and $e_i$ are charged under the flavor symmetry, with charges $[q_1]=-[e_1]=2$, $[q_2]=-[e_2]=1$. Hence, one can always perform a set of unitary rotations of $u_i,d_i,\ell_i$ such that the SM Yukawas read, after $\Phi$ condensation,
\begin{align}
    Y_d =
    \begin{pmatrix}
        z_{d_1} \eps^2 & z_{d_2} \eps^2 & z_{d_3} \eps^2 \\
        0 & y_{d_2} \eps & y_{d_3} \eps \\
        0 & 0 & x_{d_3}
    \end{pmatrix}, \quad
    Y_u = 
    \begin{pmatrix}
        z_{u_1} \eps^2 & z_{u_2} \eps^2 & z_{u_3} \eps^2 \\
        0 & y_{u_2} \eps & y_{u_3} \eps \\
        0 & 0 & x_{u_3}
    \end{pmatrix}, \quad
    Y_e = 
    \begin{pmatrix}
        z_{e_1} \eps^2 & 0 & 0 \\
        z_{e_2} \eps^2 & y_{e_2} \eps & 0 \\
        z_{e_3} \eps^2& y_{e_3} \eps & x_{e_3}
    \end{pmatrix},
    \label{eq:app1:SMYukawas}
\end{align}
with $\eps \equiv \vphi/\sqrt{2}M$. The parameters in Eq.~\eqref{eq:app1:SMYukawas} correspond to $x_{ij} ^f$ in Eq.~\eqref{eq:s2:YukLagr} and exploiting the residual rephasing symmetry of the SM fields can all be made real, except five of them which we take to be $z_{d_3}, z_{u_2}, z_{u_3}, y_{u_3}$ and $z_{e_3}$. A perturbative singular value decomposition of Eq.~\eqref{eq:app1:SMYukawas} reveals that the SM Yukawa matrices in the mass basis are simply given by the diagonal elements, 
\begin{align}
    \begin{aligned}
    (y_d, y_s, y_b) & \simeq ( z_{d_1}\eps^2,  y_{d_2}\eps, x_{d_3}),\\
    (y_u, y_c, y_t) & \simeq ( z_{u_1}\eps^2,  y_{u_2}\eps, x_{u_3}),\\
    (y_e, y_\mu, y_\tau) & \simeq ( z_{e_1}\eps^2,  y_{e_2}\eps, x_{e_3}),
    \end{aligned} 
    \label{eq:app1:SMmasses}
\end{align}
with the leading rotation matrices
\begin{align}
    \begin{gathered}
    U ^d _L \simeq  
    \begin{pmatrix}
        1 & \frac{z_{d_2}}{y_{d_2}}\eps  & \frac{z_{d_3}}{x_{d_3}} \eps^2\\
        - \frac{z_{d_2}  }{y_{d_2} }\eps & 1 & \frac{y_{d_3}}{x_{d_3}} \eps \\
        \frac{y_{d_3}  z_{d_2} - y_{d_2} z_{d_3}^*}{x_{d_3} y_{d_2}} \eps^2 & - \frac{y_{d_3}}{x_{d_3}}  \eps & 1
    \end{pmatrix}
    \simeq
    \begin{pmatrix}
        1 & \frac{m_d}{m_s} \frac{z_{d_2}}{z_{d_1}} & \frac{m_d}{m_b} \frac{z_{d_3}}{z_{d_1}} \\
        - \frac{m_d}{m_s} \frac{z_{d_2}}{z_{d_1}} & 1 & \frac{m_s}{m_b} \frac{y_{d_3}}{ y_{d_2}} \\
        \frac{m_d}{m_b} \frac{y_{d_3} z_{d_2}- y_{d_2} z_{d_3}^*}{y_{d_2} z_{d_1}} & - \frac{m_s}{m_b} \frac{y_{d_3}}{ y_{d_2}} & 1
    \end{pmatrix},
    \end{gathered}
    \label{eq:app1:CKM}
\end{align}
and similarly for $U^u _L$ and $(U^e _R)^*$. The Eq.~\eqref{eq:app1:CKM} shows that for $\Oone$ parameters, the CKM mixings can be reproduced quite well, and clarifies why in Sec.~\ref{sec:Z4} we assumed $\CKM\simeq U^d _L$: the bigger hierarchies in the up sector suppress the off-diagonal entries of $U^u _L$ to the point that a significant fine-tuning would be required for it to actually contribute to $\CKM$.

Eqs.~\eqref{eq:app1:SMmasses} and \eqref{eq:app1:CKM} can be used to fix some of the parameters in Eq.~\eqref{eq:app1:SMYukawas}, including $\eps$. The optimal benchmark, where the parameters are close to $\Oone$ and which is used to derive the bounds in the main text, is achieved for $\eps \approx 4.4 \times 10^{-3}$ with
\begin{align}
\begin{gathered}
    Y_d \simeq
    \begin{pmatrix}
        0.55     \eps^2 & 2.5 \eps^2 & (0.73 - 1.8i) \eps^2 \\
        0 & 0.049 \eps & 0.10 \eps \\
        0 & 0 & 0.011
    \end{pmatrix}, \qquad 
    Y_u \simeq 
    \begin{pmatrix}
        0.25 \eps^2 & z_{u_2} \eps^2 & z_{u_3} \eps^2 \\
        0 & 0.57 \eps & y_{u_3} \eps \\
        0 & 0 & 0.71
    \end{pmatrix},\\
    Y_e \simeq 
    \begin{pmatrix}
        0.15 \eps^2 &0 & 0 \\
         z_{e_2} \eps^2 & 0.14 \eps & 0 \\
         z_{e_3} \eps^2  & y_{e_3} \eps & 0.01
    \end{pmatrix}.
\end{gathered} 
    \label{eq:app1:YukBench}
\end{align}
where for the fit, we used the value of the SM Yukawas at 100 TeV from \cite{Huang:2020hdv}. The presence of some $\mathcal{O}(10^{-1})$ hierarchies in the lighter generations is not a problem, as it can naturally be explained by a product of several UV couplings of order $\mathcal{O}(0.3)$. Conversely, the small values of the bottom and tau Yukawas, approximately $\mathcal{O}(10^{-2})$, as indicated by $x_{d_3}$ and $x_{e_3}$, cannot be accounted for within this model and must be accepted as given.

\subsubsection{UV}
\label{sec:appUV}

Integrating out the VLF in Eq.~\eqref{eq:s3:L_VLFs} at tree-level leads to a FN EFT with Yukawas
\begin{equation}
\label{eq:app1:YukGeneric}
    Y^{(u,d)} = \begin{pmatrix}
    \epsilon_Q^2 r_Q x_1^q x_{12}^{qa} \bm{y}^{(u,d)a} \\
    \epsilon_Q x_2^{qa} \bm{y}^{(u,d)a} \\
    \bm{z}^{(u,d)}
    \end{pmatrix} \,, \qquad
    Y^{e} = \begin{pmatrix}
        \epsilon_E^2 r_E x_1^e x_{12}^{ea} \bm{y}^{ea} ~~
        \epsilon_E x_2^{ea} \bm{y}^{ea} ~~
        \bm{z}^{e \, T}
        \end{pmatrix},
\end{equation}
where $\eps_Q \equiv \vphi/\sqrt2 M_{Q_2},\eps_E \equiv \vphi/\sqrt2 M_{E_2}$. For simplicity, we have assumed degenerate $Q_2 ^a$ and $E_2 ^a$ masses and $r_Q = M_{Q_2}/M_{Q_1}$, $r_E = M_{E_2}/M_{E_1}$. Boldface quantities denote vectors in flavor space. To match Eq.~\eqref{eq:app1:YukGeneric} to \eqref{eq:app1:SMYukawas} and ultimately the SM masses and mixings, it is convenient to pick a basis in the UV theory, which simplifies calculation as much as possible. Rotation of the fields in Eq.~\eqref{eq:s3:L_VLFs} allows us to put the couplings in the following schematic form
\begin{align}
\begin{gathered}
    M_{Q_1} \sim \mathds{R}, \quad M_{Q_2} \sim \text{diag}(\mathds{R},\mathds{R}), \quad M_{E_1} \sim \mathds{R}, \quad M_{E_2} \sim \text{diag}(\mathds{R},\mathds{R}), \\ 
    {\bm z}_d \sim (0,0,\mathds{R}), \quad {\bm z}_u \sim (0,0,\mathds{R}), \quad {\bm z}_e \sim (0,0,\mathds{R}),\\
    {\bm y} ^{da} \sim 
    \begin{pmatrix}
        \mathds{R} & \mathds{C} & \mathds{C}\\
        0 & \mathds{R} & \mathds{R}
    \end{pmatrix},
    \quad
    {\bm y} ^{ua} \sim 
    \begin{pmatrix}
        \mathds{R} & \mathds{C} & \mathds{C}\\
        0 & \mathds{R} & \mathds{C}
    \end{pmatrix},
    \quad 
    {\bm y} ^{ea} \sim 
    \begin{pmatrix}
        0 & \mathds{R}\\
        \mathds{R} & \mathds{C}\\
        \mathds{R} & \mathds{C}
    \end{pmatrix},\\
    x_2 ^{qa} \sim (0, \mathds{R}), \quad x_{12} ^{qa} \sim (\mathds{R}, \mathds{C}), \quad x_1 ^q \sim \mathds{R}, \\
    x_2 ^{ea} \sim (\mathds{R},0), \quad x_{12} ^{ea} \sim (\mathds{R}, \mathds{R}), \quad x_1 ^e \sim \mathds{R},
    \end{gathered}
    \label{eq:app1:UVbasis}
\end{align}
which leads to Yukawas with the triangular structure of \eqref{eq:app1:SMYukawas}. Unfortunately, the matching is still not straightforward as not all the phases in \eqref{eq:app1:UVbasis} are physical in the EFT. A spurious phase in $\delta^d= \arg (Y_d)_{12}$ must be removed with the rotation $d_{2,3} \rightarrow e^{-i\delta^d} d_{2,3}, \bar q_{2,3} \rightarrow e^{i\delta^d} \bar q_{2,3}, u_{2,3} \rightarrow e^{-i\delta^d} u_{2,3}$ and must be taken into account when matching the UV parameters to \eqref{eq:app1:SMmasses}, \eqref{eq:app1:CKM}. Note that this rotation also affects the off-diagonal entries of $Y_u$ (unfixed in the EFT) and the Wilson coefficient of higher-dimensional operators in a non-trivial manner. Similarly, a spurious phase $\delta^e$ in $(Y_e)_{21}$ also appears, which can be removed with $ \bar \ell_{2,3} \rightarrow \bar \ell_{2,3} e^{-i \delta^e}, e_{2,3} \rightarrow e_{2,3} e^{i\delta^e}$.

The matching conditions for the down and quarks and the leptons read:
{
\allowdisplaybreaks
\begin{align}
    &\begin{cases}
        r_Q x_1 ^q x_{12} ^{q1} y^{d1}_1 = z_{d_1} = \hat y_d/\eps^2 \\
        x_2 ^{q2} y_2 ^{d2} = y_{d_2} = \hat y_s/\eps \\
        x_2 ^{q2} y_3 ^{d2} = y_{d_3} = A \lambda_c ^2 \hat y_b /\eps \\
        r_Q x_1 ^q [x_{12}^{q1} y^{d1}_2 + x_{12}^{q2} y_2 ^{d2}] = \rho_d e^{i\delta_d}, \qquad \rho_d = z_{d_2} =   \lambda_c y_{d_2}/\eps = \lambda_c \hat y_s / \eps^2\\
        r_Q x_1 ^q [x_{12}^{q1} y^{d1}_3 + x_{12}^{q2} y_3 ^{d2}]  e^{-i\delta_d} = z_{d_3} = A\lambda_c ^3 (\rho - i \eta) \hat y_b / \eps^2
    \end{cases} \qquad &(\text{down quarks}),
    \label{eq:app1:downsystem}\\
    &\begin{cases}
        r_Q x_1 ^q x_{12} ^{q1} y^{u1}_1 = z_{u_1} = \hat y_u/\eps^2 \\
        x_2 ^{q2} y_2 ^{u2} = y_{u_2} = \hat y_c/\eps \\
        x_2 ^{q2} y_3 ^{u2} = y_{u_3} \\
        r_Q x_1 ^q [x_{12}^{q1} y^{u1}_2 + x_{12}^{q2} y_2 ^{u2}] e^{-i\delta^d} = z_{u_2}\\
        r_Q x_1 ^q [x_{12}^{q1} y^{u1}_3 + x_{12}^{q2} y_3 ^{u2}] e^{-i\delta^d} = z_{u_3}
    \end{cases} \qquad &(\text{up quarks}), \\
    &\begin{cases}
        r_E x_1 ^e x_{12} ^{e2} y^{e2}_1 = z_{e_1} = \hat y_e/\eps^2 \\
        x_2 ^{e1} y_2 ^{e1} = y_{e_2} = \hat y_\mu/\eps \\
        x_2 ^{e1} y_3 ^{e1} = y_{e_3} \\
        r_E x_1 ^e [x_{12}^{e2} y^{e2}_2 + x_{12}^{e1} y_2 ^{e1}]  = \rho_{e} e^{i\delta^e}, \qquad \rho_e = z_{e_2}\\
        r_E x_1 ^e [x_{12}^{e2} y^{e2}_3 + x_{12}^{e1} y_3 ^{e1}] e^{-i\delta^e} = z_{e_3}
    \end{cases} \qquad &(\text{leptons}).
\end{align}
}
The solution to this system of equations is clearly not unique. Unfortunately, we were not able to completely solve this system analytically due to the last two equations in \eqref{eq:app1:downsystem}, which involve non-polynomial functions of the UV parameters and can be expressed as 
\begin{align}
\begin{cases}
    |x_2 + z| = b\\
    |x_3 +a z | = d\\
    \arg[x_3+az]-\arg[x_2+z] = c
\end{cases}
\label{eq:app1:newVarsSystem}
\end{align}
where $a=A\lambda_c^2 \hat y_b /\hat y_s,  b=\lambda_c \hat y_s/ \hat y_d, c=\arg A \lambda_c^3 (\rho - i\eta)$ and  $d = |A \lambda_c^3 (\rho - i\eta)| \hat y_b / \hat y_d$, and we introduced the combinations $x_2= y^{d1}_2/y^{d1}_1, x_3 = y^{d1}_3/y^{d1}_1, z = x_{12}^{q2} y^{d2}_2/( x_{12}^{q1} y^{d1}_1)$. To derive the bounds in Section~\ref{sec:Z4}, we pick a representative benchmark obtained by numerically solving \eqref{eq:app1:newVarsSystem} (again, solutions are not unique) and choosing the remaining parameters in such a way to reproduce \eqref{eq:app1:YukBench}. This reads for the quarks
\begin{align}
\label{eq:app1:UVbenchQuarks}
\begin{gathered}
{\bm z}_d = (0,0,y_b), \quad {\bm z}_d = (0,0,y_t), \\
    {\bm y} ^{da} = 
    \begin{pmatrix}
        0.26 & -0.28 - 0.96i&-0.95 +0.24i\\
        0 & 0.10 & 0.21
    \end{pmatrix},
    \quad
    {\bm y} ^{ua} = 
    \begin{pmatrix}
        0.12 & -0.57-0.20i& 0.24+0.16i\\
        0 &1.2 & -0.86+0.14i
    \end{pmatrix},\\
    r_Q = 4, \quad x_1 ^q = 0.96, \quad  x_2 ^{qa} = \left(0,0.47 \right), \quad x_{12} ^{qa} = \left(0.55, 0.07-0.99i\right), 
    \end{gathered}
\end{align}
and for the leptons
\begin{align}
\label{eq:app1:UVbenchLeptons}
\begin{gathered}
    {\bm z}_e = (0,0,y_\tau),
    \\ {\bm y} ^{ea} = 
    \begin{pmatrix}
        0 & 0.69\\
        0.38 & 0.47+0.23i\\
        0.49 & 0.19+0.52i
    \end{pmatrix},\\
    r_E = 0.5, \quad x_2 ^{ea} = (0.39,0), \quad x_{12} ^{ea} = (0.45, 0.54), \quad x_1 ^e = 0.8.
    \end{gathered}
\end{align}
with $\delta^d = -1.2$ and $\delta^e = 0.28$. This benchmark fixes the remaining entries of \eqref{eq:app1:YukGeneric}, after the $\delta^d, \delta^u$ rotation, to $z_{e_2} = 0.18 , z_{e_3} = 0.15 + 0.07i, y_{e_3} = 0.19$ and $z_{u_2} = 2.3+0.06i, z_{u_3} = -2.1+0.56i, y_{u_3} = -0.42+0.07i$.

\subsection{$\Z_8$ model}
\label{sec:app:z8}

In this case, with flavor rotations, we can put the Yukawa matrices in the form
\begin{align}
    Y_d =
    \begin{pmatrix}
        z_{d_1} \eps^4 & z_{d_2} \eps^4 & z_{d_3} \eps^4 \\
        0 & y_{d_2} \eps^3 & y_{d_3} \eps^3 \\
        0 & 0 & x_{d_3} \eps^2
    \end{pmatrix}, \quad
    Y_u = 
    \begin{pmatrix}
        z_{u_1} \eps^4 & z_{u_2} \eps^3 & z_{u_3} \eps^2 \\
        y_{u_1} \eps^3 & y_{u_2} \eps^2 & y_{u_3} \eps \\
        x_{u_1} \eps^2 & x_{u_2} \eps & x_{u_3}
    \end{pmatrix}, \quad
    Y_e = 
    \begin{pmatrix}
        z_{e_1} \eps^4 & 0 & 0 \\
        z_{e_2} \eps^4 & y_{e_2} \eps^3 & 0 \\
        z_{e_3} \eps^4 & y_{e_3} \eps^3 & x_{e_3}\eps^2 
    \end{pmatrix},
    \label{eq:app1:SMYukawasZ8}
\end{align}
where in $Y_d$ and $Y_e$ the only complex entries are $z_{d_3}$ and $z_{e_3}$, while in $Y_u$ all entries are complex except the diagonal ones. Assuming for simplicity that the off-diagonal entries of $Y_u$ are small, the SM Yukawas in the mass basis are given as
\begin{align}
    \begin{aligned}
    (y_d, y_s, y_b) & \simeq ( z_{d_1}\eps^4,  y_{d_2}\eps^3, x_{d_3} \eps^2),\\
    (y_u, y_c, y_t) & \simeq ( z_{u_1}\eps^4,  y_{u_2}\eps^2, x_{u_3}),\\
    (y_e, y_\mu, y_\tau) & \simeq ( z_{e_1}\eps^4,  y_{e_2}\eps^3, x_{e_3}\eps^2), 
    \end{aligned}
    \label{eq:app1:SMmassesZ8}
\end{align}
with the CKM matrix still given by Eq.~\eqref{eq:app1:CKM}. We find that the best benchmark in this case is given by $\eps \simeq 6.6 \times 10^{-2}$, leading to 
\begin{align}
\begin{gathered}
    Y_d \simeq
    \begin{pmatrix}
        0.55 \eps^4 & 2.5 \eps^4 & (0.73 - 1.84i) \eps^4 \\
        0 & 0.74 \eps^3 & 1.51 \eps^3 \\
        0 & 0 & 2.4 \eps^2
    \end{pmatrix}, \quad
    Y_u \simeq
    \begin{pmatrix}
        0.25 \eps^4 & z_{u_2} \eps^3 & z_{u_3} \eps^2 \\
        y_{u_1} \eps^3 & 0.57 \eps^2 & y_{u_3} \eps \\
        x_{u_1} \eps^2 & x_{u_2} \eps & 0.71
    \end{pmatrix}, \\
    Y_e \simeq 
    \begin{pmatrix}
        0.15 \eps^4 & 0 & 0 \\
        z_{e_2} \eps^4 & 2.1 \eps^3 & 0 \\
        z_{e_3} \eps^4 & y_{e_3} \eps^3 & 2.3\eps^2 
    \end{pmatrix}.
\end{gathered}
\label{eq:app1:SMYukawasZ8num}
\end{align}
Compared to Eq.~\eqref{eq:app1:SMYukawas} the fit in this case is better in terms of the parameters being close to $\Oone$, with a maximum spread between the smallest and the biggest ones of $\mathcal{O}(10)$.

\section{Derivation of the ALP couplings}
\label{sec:app:ALP}
Assuming a generic Yukawa Lagrangian
\begin{equation}
    \mathcal{L} \supset -y_{ij}^f \left(\frac{\Phi}{M}\right)^{n_{ij}^f} \bar{F}_i H f_j + \textrm{h.c.}\,,
\end{equation}
where $F$ is a left-handed $\SU(2)_L$ doublet and $f$ is a right-handed singlet, we now compute the $\rho$ and $a$ couplings to the fermions. After SSB, we write
\begin{equation}
    \Phi = \left(\frac{v_\Phi+\rho}{\sqrt{2}}\right) e^{ia/\vphi}, \qquad 
    H = \begin{pmatrix}
         0 \\
         \frac{v_{\mathrm{EW}} + h}{\sqrt 2}
     \end{pmatrix}\,.
\end{equation}
Plugging this into the Lagrangian above and expanding the obtained expression to leading order in $\rho/v_\Phi$ and $a/v_\Phi$, we obtain
\begin{equation}
    \mathcal{L} \supset -y_{ij}^f \left(\frac{v_\Phi}{\sqrt{2} M}\right)^{n_{ij}^f}\left(1+{n_{ij}^f}\frac{\rho+i a}{v_\Phi}\right) \bar{f}_{Li} \left(\frac{v_\mathrm{EW}+h}{\sqrt{2}}\right)f_{Rj} + \textrm{h.c.} + \textrm{h.o.}\,.
\end{equation}
where $f_L$ is the appropriate component from the doublet $F$. For up-type quarks, $\tilde{H}$ would be used instead. We define the FN expansion parameter and the fermion mass matrices
\begin{equation}
    \epsilon = \frac{v_\Phi}{\sqrt{2}M} \,, \qquad m_{ij}^f =  \frac{v_\mathrm{EW}}{\sqrt{2}} y_{ij}^f \epsilon^{n_{ij}^f}\,,
\end{equation}
seeing the FN mechanism at work once again. With this, we can write the Lagrangian as
\begin{equation}
    \mathcal{L} \supset - \sum_{f=u,d,e} \left[m_{ij}^{f} \bar{f}_{Li}\left(1+\frac{h}{v_\mathrm{EW}}\right)f_{Rj} + \frac{m_{ij}^{f} n_{ij}^{f}}{v_\Phi} \bar{f}_{Li} \left(\rho+ia\right) f_{Rj}\right] + \textrm{h.c.} \,.
\end{equation}
Next, the fermion masses should be diagonalized as usual
\begin{equation}
f_{R} \rightarrow U_{R}^{f} f_{R}\,, \quad f_{L} \rightarrow U_{L}^{f} f_{L}\,, \quad U_L^{f\dagger} m^f U_R^f = \hat{m}^f\,,
\end{equation}
where $\hat{m}^f$ is a diagonal matrix, $\hat{m}_{ij}^f\equiv m_i^f \delta_{ij}$. With this, the Higgs-Yukawa couplings are also diagonalized, whereas the $a$ and $\rho$ couplings are not. This induces, in general, flavor-changing neutral currents. We then have in the fermion mass basis
\begin{equation}
    \mathcal{L} \supset - \sum_{f=u,d,e} \left[m_{i}^{f} \bar{f}_{Li}\left(1+\frac{h}{v_\mathrm{EW}}\right)f_{Ri} + c_{ij}^f \bar{f}_{Li} \left(\rho + i a \right) f_{Rj}\right] + \textrm{h.c.}\,.
\end{equation}
with
\begin{equation}
    c_{ij}^f = \frac{1}{v_\Phi}(U_L^{f\dagger})_{ia} m_{ab}^f n_{ab}^f U_{Rbj}^{f}\,.
\end{equation}
We can further rewrite this by noting that $n_{ab}^f = [F_a] - [f_b]$ is a matrix built from (integer) FN charges. By defining diagonal charge matrices $Q^f_{ij} \equiv [f_i] \delta_{ij}$, the combination of matrices appearing in the middle can be expressed as
\begin{equation}
    m_{ab}^{f} n_{ab}^{f} = (Q^F m^f - m^f Q^f)_{ab} \,.
\end{equation}
With this, the $c^f$ coupling matrices can be expressed as
\begin{equation}
\begin{split}
    c^f &= \frac{1}{v_\Phi} \left[ U_L^{f\dagger} \left(Q^F m^f - m^f Q^f\right) U_R^f \right] \\
    &= \frac{1}{v_\Phi} \left[ U_L^{f\dagger} Q^F\left(U_L^f U_L^{f\dagger}\right) m^f U_R^f - U_L^{f\dagger} m^f \left(U_R^f U_R^{f\dagger}\right) Q^f U_R^f \right] \\
    &= \frac{1}{v_\Phi }\left[ \left(U_L^{f\dagger} Q^F U_L^f \right) \hat{m}^f-  \hat{m}^f \left(U_R^{f\dagger} Q^f U_R^f \right)\right]~.
\end{split}
\label{eq:appB:cformula}
\end{equation}
It is worth reemphasizing that this is now expressed in terms of diagonal fermion mass matrices $\hat{m}^f$ and diagonal FN charge matrices $Q^F$ and $Q^f$ for the $\SU(2)$ doublet and singlet, respectively.


\bibliographystyle{JHEP}
\bibliography{biblio.bib}

\end{document}